\documentclass[a4paper]{article}
\usepackage{graphicx}

\begin{document}

\begin{titlepage}
\title{\Large \bf A Guide to Modeling Credit Term Structures\footnote{This review is based in part on the series of papers which the author (co)wrote several years ago \cite{BMW-1,BMW-2,BMW-3,Berd-ForwardCDS,Berd-RecoverySwaps}. To appear in {\em Oxford Handbook of Credit Derivatives}, eds.\ A. Lipton and A. Rennie, Oxford University Press, 2010.}}
\author{{\large \bf Arthur M. Berd} \\
Capital Fund Management \\
6, boulevard Haussmann, 75009 Paris, France}
\date{December 17, 2009}
\maketitle

\begin{abstract}
We give a comprehensive review of credit term structure modeling methodologies. The conventional approach to modeling credit term structure is summarized and shown to be equivalent to a particular type of the reduced form credit risk model, the fractional recovery of market value approach. We argue that the corporate practice and market observations do not support this approach. The more appropriate assumption is the fractional recovery of par, which explicitly violates the strippable cash flow valuation assumption that is necessary for the conventional credit term structure definitions to hold. We formulate the survival-based valuation methodology and give alternative specifications for various credit term structures that are consistent with market observations, and show how they can be empirically estimated from the observable prices. We rederive the credit triangle relationship by considering the replication of recovery swaps. We complete the exposition by presenting a consistent measure of CDS-Bond basis and demonstrate its relation to a static hedging strategy, which remains valid for non-par bonds and non-flat term structures of interest rates and credit risk.
\end{abstract}

\end{titlepage}

\tableofcontents

\newpage

\section{Introduction} \label{sec:introduction}

Most of fixed income valuation and risk methodologies are centered on modeling yield and spread term structures. The main reason for this is that the vast majority of debt instruments exhibit very high correlation of price returns. Therefore the common pricing factors encoded in the yield curve have a high explanatory power. This is especially true for Treasury bonds, where the market is extremely efficient and any deviation of individual bond valuation from the common curve is quickly arbitraged away. 

For corporate bonds the common yield curves are much less binding, since the market is substantially less liquid, even for investment grade benchmark issuers. The other driving factors of the valuation of credit-risky bonds are the credit quality of the name, the estimated recovery in case of default, geographic and industry as well as the issuer and security specifics.

The standard market practice in analyzing investment grade credit bonds is to introduce the notion of spread to base (Treasury or swaps) curve. The sector or issuer spread curves are thought to reflect the additional specific information besides the underlying base yield curve. Among many definitions of spread measures used by practitioners who analyze credit bonds the most robust and consistent one is the OAS, also known as the Z-spread in case of bonds with no embedded options (in the sequel we use these terms interchangeably.)

While the Z-spread term structures are commonly used to quote and analyze relative value among credit bonds, it is well known that these measures become inadequate for distressed credits for which the market convention reverts to quoting bond prices rather than spreads. This market convention reflects a fundamental flaw in the conventional credit valuation methodology, namely the assumption that a credit bond can be priced as a portfolio of cash flows discounted at a risky rate. The reason why this assumption is wrong is that the credit-risky bonds do not have fixed cash flows, and in fact their cash flows are variable and dependent on the future default risk even if contractually fixed. This methodological breakdown holds for any credits, whether investment grade, or distressed, with the errors being relatively small for investment grade bonds trading near par but growing rapidly with the rise in default risk expectations. 

We develop an alternative valuation methodology and introduce new definitions of credit term structures that are consistent with the valuation of bonds for all credit risk levels. We propose an efficient and robust empirical estimation procedure for these new measures of credit risk and show examples of fitted credit term structures for both low and high default risk cases. The resulting relative value measures may differ from conventional ones not only quantitatively but also qualitatively - possibly even reversing the sign of rich/cheap signals and altering the composition of relative value trades.

The current methodology has been implemented within Lehman Brothers (now Barclays Capital) quantitative credit toolkit since 2003 \cite{BMW-1, BMW-2, BMW-3} and has been used by many investors focusing both on cash bonds and credit derivatives since then. It successfully stood the test of extreme stress during the recent credit crisis, while the conventional credit measures have become unusable. We hope that this experience will lead to the proposed methodology gaining in acceptance and that this review will help investors, traders and researchers to become familiar with it and use it in daily practice.

\section{The Conventional Term Structure Measures}  \label{sec:conventional}

In this Section we re-examine critically the standard lore of credit bond valuation methodologies in order to understand better their breakdown in case of distressed bonds which we mentioned in the Introduction. For a more detailed reference on the standard fixed income valuation methodology see \cite{Tuckman-book}.

\subsection{The Strippable Cash Flow Valuation Methodology} \label{sec:strippable_methodology}

The main assumption of the strippable cash flow valuation methodology is that fixed-coupon debt instruments can be priced as a portfolio of $N$ individual cash flows $CF(t_{i})$ using the discount function $Z(t_{i})$, which coincides with the present value of a zero coupon bond maturing at some future time $t_{i}$ as observed today. This conclusion follows immediately from the assumption that the cash flows are completely fixed in all future states of the world, and therefore a portfolio of zero coupon bonds with maturities chosen to match the contractual cash flow payment dates and the notional chosen to equal the cash flow payment amounts will indeed provide a complete replication of the bond in all future states of the world, and therefore the present value of this portfolio should coincide with the present value of the bond:

\begin{eqnarray} \label{eq:discounted_cf}
PV_{bond} & = & PV\left\{ \sum_{i=1}^{N} CF(t_{i}) \right\} \nonumber \\ 
 & = & \sum_{i=1}^{N} CF(t_{i}) \, PV\left\{ 1(t_{i}) \right\}  \nonumber \\
 & = & \sum_{i=1}^{N} CF(t_{i}) \, Z(t_{i})
\end{eqnarray}

Such an assumption is clearly valid for default risk-free instruments such as U.S. Treasury bonds which can be contractually stripped, separating their coupon and notional cash flows. The Treasury strips trade in a liquid market and are often referenced for an estimation of the fair value discount term structure. 

The strippable cash flow methodology is commonly extended to credit-risky bonds by assuming that they can be priced using a similarly defined ``risky discount function'' $Z_{risky}(T)$. Since the pricing equation (\ref{eq:discounted_cf}) is deterministic, one can express the risky discount function without loss of generality as a product of the risk-free base discount function and the risky excess discount function:

\begin{equation} \label{eq:risky_discfunc}
Z_{risky}(t) = Z_{base}(t) \, Z_{excess}(t)
\end{equation}

Thus, in the conventional strippable cash flow methodology the fundamental pricing equation for a credit risky bond reads as follows:

\begin{equation} \label{eq:discounted_cf_risky}
PV_{bond} = \sum_{i=1}^{N} CF(t_{i}) \, Z_{base}(t_{i}) \, Z_{excess}(t_{i})
\end{equation}

Let us express the same in plain language: {\em The present value of a contractually-fixed cash flow security under a strippable cash flows valuation framework is equal to the sum of the present values of the individual cash flows.} 

In other words, the conventional valuation methodology hinges on the ability to represent a fixed income security as a portfolio of individual cash flows. Whether or not such a representation is possible for credit bonds, and what discount function applies if it is, depends on the realities of the market, and will be discussed in Section \ref{sec:phenomenology}. 

\subsection{The Conventional Bond Yield and Spread Measures} \label{sec:conventional_term_structure}

In conventional approach the modeling of credit bonds centers on defining various spread measures to relevant base yield curves, which can be the Treasury curve or the swaps curve. We give only a cursory definition of the base curve for completeness of exposition, and refer the reader to \cite{Tuckman-book} for details on the methodologies for defining and estimating base curve, and to \cite{OKane-Sen-2005} for conventional spread measures.

\subsubsection{Risk-Free Bonds}  \label{sec:risk_free_bonds}

Before proceeding to the case of credit-risky bonds, let us first define the term structure metrics used for pricing risk-free bonds. A convenient parameterization of the base discount function $Z(t)$ is given by the term structure of zero-coupon interest rates $r(t)$ in a continuous compounding convention:

\begin{equation} \label{eq:zc_yield_curve}
Z(t) = e^{-r(t) \, t}
\end{equation}

In this definition, the interest rates $r(t)$ correspond to pricing of zero-coupon risk-free bonds, as can be seen from (\ref{eq:discounted_cf}), for which there is only a single unit cash flow at time $t$. One could express these rates in some fractional compounding fequency, such as annual or semi-annual, however given that there is no particular frequency tied to the coupon payments, it is more convenient to use the continuous compounding as in (\ref{eq:zc_yield_curve}).

In a dynamic context, one can define a stochastic process for the short rate $r_{t}$, such that the expected discount function $Z(T)$ is the expected value of the unit amount payable at time $T$, with stochastic discount $Z_{t}$ given by the compounded short-term rates along each possible future realization:

\begin{equation} \label{eq:expected_zc_discount}
Z(t) = E_{t=0}\left\{ Z_{t} \right\} = E_{t=0}\left\{ e^{- \int_{0}^{t} r_{s} \, ds} \right\}
\end{equation}

Note the notational difference between $r(t)$ and $Z(t)$ which are deterministic functions of a finite maturity $t$ measured at the initial time, and the stochastic processes $r_{t}$ and $Z_{t}$ denoting the value of the random variable at future time $t$. We will use the subscripts to denote the time dependence of stochastic processes, and function notation for deterministic variables of term to maturity.

A useful counterpart of the interest rate for finite maturity $t$ is the forward rate which is defined as the breakeven future discounting rate between time $t$ and $t+dt$. 

\begin{equation} 
\frac{Z(t+dt)}{Z(t)} = e^{-f(t) \, dt} \label{eq:fwd_rate_def}
\end{equation}

The relationships between the discount function, forward rate and zero-coupon interest rate (in continuous compounding) can be summarized as follows:

\begin{eqnarray} 
f(t) & = & -\frac{\partial}{\partial t} \log Z(t) \label{eq:fwd_rate} \\
r(t) & = & -\frac{1}{t} \int_{0}^{t} f(s) ds \label{eq:zero_vs_fwd_rate}
\end{eqnarray}

One can follow the celebrated Heath, Jarrow and Morton \cite{HJM} approach to define the stochastic process for forward rates, and use the identity $r_{t} = f_{t}(t)$ to derive the expected discount function as in (\ref{eq:expected_zc_discount}). While this is certainly one of the mainstream approaches in modeling risk-free interest rates, it is less often used to model credit and we will not discuss it in length in this chapter.

Another often used metric is the yield to maturity $y(t)$. In a compounding convention with frequency $q$ periods per annum, the standard pricing equation for a fixed-coupon bond is:

\begin{equation} \label{eq:yield_to_maturity}
PV_{bond} = \sum_{i=1}^{N} \frac{C/q}{\left(1+y/q\right)^{q \, t_{i}}} + \frac{1}{\left(1+y/q\right)^{q \, t_{N}}}
\end{equation}

Given a present value of a bond (including its accrued coupon), one can determine the yield to maturity from (\ref{eq:yield_to_maturity}). Of course, one must remember that unlike the term structure of interest rates $r(t)$, the yield to maturity $y(T)$ depends on the specific bond with its coupon and maturity. In particular, one cannot apply the yield to maturity obtained from a 5-year bond with 4.5\% coupon to value the 5-th year cash flow of a 10-year bond with 5.5\% coupon. In other words, the yield to maturity is not a metric that is consistent with a strippable discounted cash flow valuation methodology, because it forces one to treat all the coupons and the notional payment as inseparable under the equation (\ref{eq:yield_to_maturity}). 

In contrast, the zero-coupon interest rate $r(t)$ is well suited for fitting across all cash flows of all comparable risk-free bonds, Treasury strips, etc., with a result being a universal fair value term structure that can be used to price any appropriate cash flows. Therefore, we will primarily rely in this report on the definition (\ref{eq:zc_yield_curve}) for the base discount function.

\subsubsection{Bullet Bonds: Yield Spread}  \label{sec:yield_spread}

The simplest credit-risky bonds have a `bullet' structure, with a fixed coupon and the entire notional payable at maturity. Such bonds are free of amortizing notional payments, coupon resets, and embedded options, which could complicate the cash flow picture. Fortunately for the modelers, the bullet bonds represent the majority of investment grade credit bond market and therefore their modeling is not just an academic exercize. 

The simplest (and naive) approach to valuaing a credit bond is to apply the same yield-to-maturity methodology as for the risk-free bonds (\ref{eq:yield_to_maturity}):

\begin{equation} \label{eq:yield_to_maturity_credit}
PV_{bond} = \sum_{i=1}^{N} \frac{C/q}{\left(1+Y/q\right)^{q t_{i}}} + \frac{1}{\left(1+Y/q\right)^{q t_{N}}}
\end{equation}

The corresponding risky yield to maturity $Y(T)$ is then often compared with the yield to maturity of a benchmark reference Treasury bond $y(T)$ which is usually chosen to be close to the credit bond's maturity $T=t_{N}$. The associated spread measure is called the yield spread (to maturity):

\begin{equation} \label{eq:yield_spread}
S_{Y}(T) = Y(T) - y(T)
\end{equation}

Since the benchmark Treasury bonds are often selected among the most liquid ones (recent issue 2, 5, 7, 10 or 30 year bonds), the maturity mismatch in defining the yield spread can be quite substantial. A slightly improved version of the yield spread is the so called interpolated spread, or I-spread. Instead of using a single benchmark Treasury bond, it refers to a pair of bonds whose maturities bracket the maturity of the credit bond under consideration. Suppose the index 1 refers to the Treasury bond with a shorter maturity, and the index 2 refers to the one with longer maturity. The linearly interpolated I-spread is then defined as:

\begin{equation} \label{eq:int_yield_spread}
S_{I}(T) = Y(T) - \left(\frac{T_2 - T}{T_2 - T_1} y(T_1) + \frac{T - T_1}{T_2 - T_1} y(T_2) \right)
\end{equation}

We emphasize that the yield spread (or the I-spread) measure for credit bonds calculated in such a manner is rather useless and can be very misleading. The reason is that it ignores a whole host of important aspects of bond pricing, such as the shape of the term structure of yields, the coupon and price level, the cash flow uncertainty of credit bonds, etc. We shall discuss these limitations in latter parts of this article. For now, it suffices to say that even if one ignores the intricate details of credit risk, the yield spread is still not a good measure, and can be substantially improved upon.

\subsubsection{Bullet Bonds: Z-Spread}  \label{sec:z_spread}

As we noted in eq.\ (\ref{eq:discounted_cf_risky}), under the assumption of strippable cash flows, the pricing adjustment for a credit bond is contained within an excess discount function. This function can be expressed in terms of the Z-spread $S_{Z}(t)$:

\begin{equation} \label{eq:z_spread}
Z_{excess}(t) = e^{-S_{Z}(t) \, t}
\end{equation}

Note, that unlike the yield and spread to maturity, the Z-spread is tied only to the discount function, which in turn is assumed to be a universal measure for valuaing all cash flows with similar maturity for the same credit, not just those of the particular bond under the consideration. The corresponding risky discounting rate is consequently defined as:

\begin{equation} \label{eq:risky_rate}
R(t) = r(t) + S_{Z}(t)
\end{equation}
such that the total risky discount function is related to the risky yield in the same manner as the risk-free discount function is to risk-free zero-coupon yield, and the strippable cash flow valuation framework (\ref{eq:discounted_cf_risky}) is assumed to hold:

\begin{equation} \label{eq:zc_risky_yield_curve}
Z_{risky}(t) = e^{-R(t) \, t}
\end{equation}

As we shall argue in the Section \ref{sec:phenomenology}, this assumption is, in fact, explicitly violated by actual credit market conventions and practices. However, Z-spread does remain moderately useful for high grade credit bonds which have very low projected default probability. Therefore, we encourage its use as a shortcut measure, but urge the analysts to remember about its limitations.

Similarly to the definition of the forward risk-free interest rates (\ref{eq:fwd_rate}), one can define the forward risky rates $F(t)$ and forward Z-spreads $S_{F}(t)$:

\begin{eqnarray} 
F(t) & = & -\frac{\partial}{\partial t} \log Z_{risky}(t)  \label{eq:risky_fwd_rate} \\
R(t) & = & -\frac{1}{t} \int_{0}^{t} F(s) ds \\
S_{F}(t) & = & -\frac{\partial}{\partial t} \log Z_{excess}(t)  \label{eq:risky_fwd_spread} \\
S_{Z}(t) & = & -\frac{1}{t} \int_{0}^{t} S_{F}(s) ds
\end{eqnarray}

We will see from subsequent discussion that the forward Z-spread $S_{F}(t)$ has a particular meaning in reduced form models with so called fractional recovery of market value, relating it to the hazard rate of the exogenous default process. This fundamental meaning is lost, however, under more realistic recovery assumptions.

\subsubsection{Callable/Puttable Bonds: Option-Adjusted Spread (OAS)}  \label{sec:OAS}

For bonds with embedded options, the most widely used valuation measure is the option-adjusted spread (OAS). The OAS can be precisely defined for risk-free callable/puttable bonds and allows one to disaggregate the value of an embedded option from the relative valuation of bond's cash flows compared to the bullet bond case. 

One way to define the OAS is to explicitly define the stochastic model for the short rates process $r_{t}$ to value the embedded option, and to assume that the risk-free interest rate in this model is bumped up or down by a constant amount across all future times, such that after discounting the variable bond plus option cash flows $CF_{t}$ with the total interest rate $r_{t} + OAS$ one gets the market observed present value of the bond. 

\begin{equation} \label{eq:OAS}
PV_{bond} = E_{t=0}\left\{ \sum_{t=0}^{T}{ CF_{t} \, e^{-\sum_{u=0}^{t} (r_{u} + OAS) dt} } \right\}
\end{equation}

Note that this definition does not necessarily coincide with an alternative assumption where one adds a constant $OAS$ to the initial term structure of interest rates. Whether such initial constant will translate into a constant shift of future stochastic rates $r_{0}(t)+OAS \rightarrow r_{t}+OAS$, depends on the details of the short rate process. Generally speaking, only processes which are linear in $r_{t}$ will preserve such a relationship, which frequently used log-normal or square-root processes will not.

The relationship between the OAS and Z-spread becomes clear if one assumes a zero volatility of interest rates in the equation (\ref{eq:OAS}). In this case, the evolution of the short rate $r_{t}$ becomes a deterministic function $r_{t} = f(t)$, and the random cash flows reduce to the deterministic cash flows `to worst' $CF_{w}(t)$, as only one of the many embedded options (including maturity) will be the deepest in the money. Thus, under this assumption, the OAS coincides with the Z-spread calculated for the `to worst' cash flow scenario -- the `Z' in Z-spread stands for {\em zero volatility}.

The discounted value of `to worst' cash flows reflects the intrinsic value of the embedded option. Therefore, the difference between the full $OAS$ calculated from (\ref{eq:OAS}) under stochastic interest rates and the Z-spread calculated under zero volatility assumption reflects the time value of the embedded option, plus any security-specific pricing premia, if any. 

The base curve used for both OAS and Z-spread calculation is usually chosen to be the swaps curve, such that the base rate plus OAS resemble a realistic reinvestment rate for bond cash flows. While it is possible to use any other base discount curve, such a Treasury zero-coupon rate term structure, the meaning of the OAS in those cases will be altered to include both security-specific and funding risks.

\subsubsection{Floating Rate Notes: Discount Margin}  \label{sec:libor_spreads}

The Floating Rate Notes structure greatly mitigates market risk, insulating these securities from the market-wide interest rate fluctuations. The credit risk, however, still remains and the conventional methodology applies the additional discount spread to value the variable future cash flows. These variable cash flows are typically tied to the LIBOR-based index, such as 3-month or 12-month LIBOR rate, with a contractually specified additional {\em quoted margin}:

\begin{equation} \label{eq:FRN-QM}
CF^{FRN}_{i} = L(t_{i}) + QM + 1_{\left\{t_{i}=t_{N}\right\}}
\end{equation}
where $L(t_{i})$ is the future LIBOR rate fixing valid for the period including the payment time $t_{i}$, and we have added an explicit term for the final principal payment. 

In a conventional approach, instead of the unknown future rate $L(t_{i})$ one takes the forward LIBOR rate $L(t_{i-1},t_{i})$ estimated for the reset time $t_{i-1}$ and a forward tenor $\Delta_{i} = t_{i}-t_{i-1}$ until the next payment date, and thus reduces the valuation problem to a detrministic (zero volatility) one. Having projected these deterministic future cash flows, one can then proceed to value them with a discount curve whose rate is given by the same forward LIBOR rates with an added spread amount called {\em zero discount margin}. 

\begin{eqnarray} \label{eq:FRN-DM}
PV_{bond} & = & \sum_{i=1}^{N}{ CF^{FRN}_{i} \, Z_{float}(t_{i})} \nonumber \\
Z_{float}(t_{i}) & = & \prod_{i=1}^{N}{\frac{1}{1+\Delta_{i} \, (L(t_{i-1},t_{i}) + DM)}}
\end{eqnarray}

The difference between the zero discount margin $DM$ and the quoted margin $QM$ reflects the potential non-par valuation of the credit-risky FRN which is due to changing default and recovery risks since the issuance. If the credit risk remained stable (and neglecting the possible term structure effects), the zero discount margin would be identical to the original issue quoted marging, and the FRN would be valued at par on the corresponding reset date. Unlike this, in case of a fixed coupon bond the price can differ from par even if the credit risk remained unchanged, solely due to interest rate changes or yield curve roll-down. 

\section{The Phenomenology of Credit Pricing} \label{sec:phenomenology}

\subsection{Credit Bond Cash Flows Reconsidered}

The root of the problem with the conventional strippable cash flow methodology as applied to credit-risky bonds is that credit bonds do not have fixed cash flows. Indeed, the main difference between a credit risky bond and a credit risk-free one is precisely the possibility that the issuer might default and not honor the contractual cash flows of the bond. In this event, even if the contractual cash flows were fixed, the realized cash flows may be very different from the promised ones. 

Once we realize this fundamental fact, it becomes clear that the validity of the representation of a credit bond as a portfolio of cash flows critically depends on our assumption of the cash flows in case of default. In reality, when an issuer defaults, it enters into often protracted bankruptcy proceedings during which various creditors including bondholders, bank lenders, and those with trading and other claims on the assets of the company settle with the trustees of the company and the bankruptcy court judge the priority of payments and manner in which those payments are to be obtained. 

Of course, modeling such an idiosyncratic process is hopelessly beyond our reach. Fortunately, however, this is not necessary. Once an issuer defaults or declares bankruptcy its bonds trade in a fairly efficient distressed market and quickly settle at what the investors expect is the fair value of the possible recovery. 

The efficiency of the distressed market and the accuracy and speed with which it zooms in on the recovery value is particularly high in the U.S., as evidenced by many studies (see \cite{LossCalc} and references therein). Assuming that the price of the bond immediately after default represents the fair value of the subsequent ultimate recovery cash flows, we can simply take that price as the single post-recovery cash flow which substitutes all remaining contractual cash flows of the bond. 

The market practice is to use the price approximately one month after the credit event to allow for a period during which the investors find out the extent of the issuer's outstanding liabilities and remaining assets. This practice is also consistent with the conventions of the credit derivatives market, where the recovery value for purposes of cash settlement of CDS is obtained by a dealer poll within approximately one month after the credit event.

From a modeling perspective, using the recovery value after default as a replacement cash flow scenario is a well established approach. However, the specific assumptions about the recovery value itself differ among both academics and practitioners. We will discuss these in detail in the next section as we develop a valuation framework for credit bonds. But first we would like to explore a few general aspects of credit bond valuation that will set the stage for an intuitive understanding of the formal methodology we present later in this chapter.

\subsection{The Implications of Risky Cash Flows}

The relevance of the uncertain cash flows of credit-risky bonds depends on the likelihood of the scenarios under which the fixed contractual cash flows may altered. In other words, it depends on the level of default probability\footnote{Strictly speaking, one must also consider the possibility of debt restructuring as another scenario where the contractually fixed cash flows will be altered. Assuming that such restructuring is done in a manner which does not change the present value of the debt (otherwise either the debtors or the company will prefer the bankruptcy option) one can argue that its effect on valuing credit bonds should be negligible. Of course, in practice the restructurings are often done in a situation where either the bondholders or the company is in a weak negotiating position, thereby leading to a non-trivial transfer of value in the process. We shall ignore this possibility in the present study.}. As we will find out shortly, what one has to consider here is not the real-world (forecasted) default probability, but the so-called implied (or breakeven) default probability.

One of the most striking examples of the mis-characterization of credit bonds by the conventional spread measures is the often cited steeply inverted spread curve for distressed bonds. One frequently hears explanations for this phenomenon based on the belief that the near-term risks in distressed issuers are greater than the longer term ones which is supposedly the reason why the near-term spreads are higher than the long maturity ones. However, upon closer examination one can see that the inverted spread curve is largely an `optical' phenomenon due to a chosen definition of the spread measure such as the Z-spread rather than a reflection of the inherent risks and returns of the issuer's securities.

Indeed, the more important phenomenon in reality is that once the perceived credit risk of an issuer become high enough, the market begins to price the default scenario. In particular, investors recognize what is known in the distressed debt markets as the {\em acceleration of debt} clause in case of default. The legal covenants on most traded bonds are such that, regardless of the initial maturity of the bond, if the issuer defaults on any of its debt obligations, all of the outstanding debt becomes due immediately. This is an important market practice designed to make sure that, in case of bankruptcy, investor interests can be pooled together by their seniority class for the purposes of the bankruptcy resolution process. As a result, both short and long maturity bonds begin trading at similar dollar prices - leading to a flat term structure of prices. 

Let us now analyze how this translates into a term structure of Z-spreads. In the conventional spread-discount based methodology, one can explain an \$80 price of a 20-year bond with a spread of 500bp. However, to explain an \$80 price for a 5-year bond, one would need to raise the spread to very large levels in order to achieve the required discounting effect. The resulting term structure of spreads becomes downward sloping, or inverted. The inversion of the spread curve is due to the bonds trading on price, which is strongly dependent on the expected recovery, while the Z-spread methodology does not take recoveries into account at all! 

In the survival-based methodology, which we describe in this chapter, the low prices of bonds are explained by high default rates, which need not have an inverted term structure. A flat or even an upward sloping term structure of default rates can lead to an inverted Z-spread term structure if the level of the default rates is high enough. This is not to say that the near-term perceived credit risks are never higher than the longer term ones - just that one cannot make such a conclusion based on the inverted Z-spread term structure alone.

\begin{figure}[t]
\includegraphics[height=3in,width=4.5in]{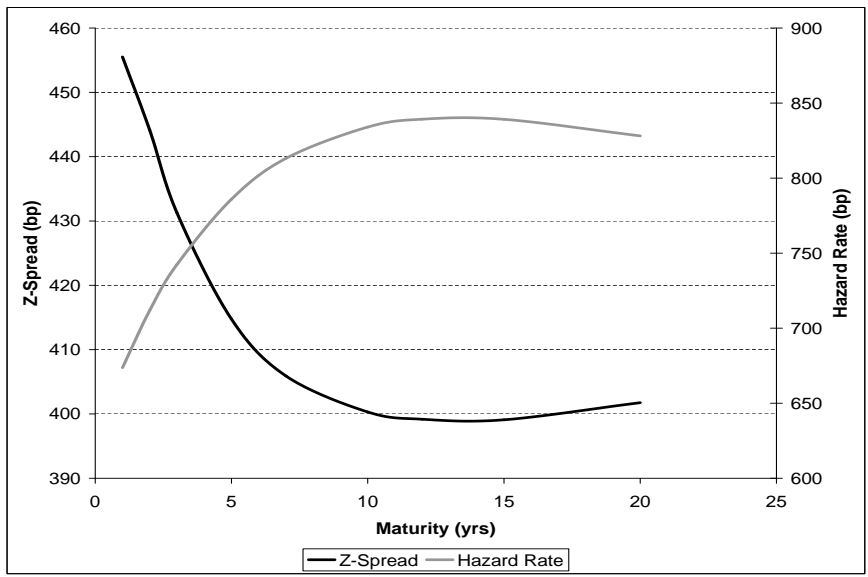}%
\caption{{\small Optically inverted Z-spread term structure. Ford Motor Credit, 12/31/2002.}}%
\label{fig:optical-inversion}%
\end{figure}

Consider for example the credit term structure of Ford Motor Co. as of 12/31/2002 shown in Figure \ref{fig:optical-inversion}. The default hazard rate is fitted using the survival-based methodology developed later in this chapter, and is upward sloping with a hump at 15 year range. However, the Z-spread curve fitted using the conventional methodology which does not take into account the potential variability of cash flows, is clearly inverted. 

\begin{figure}[t]
\includegraphics[height=3in,width=4.5in]{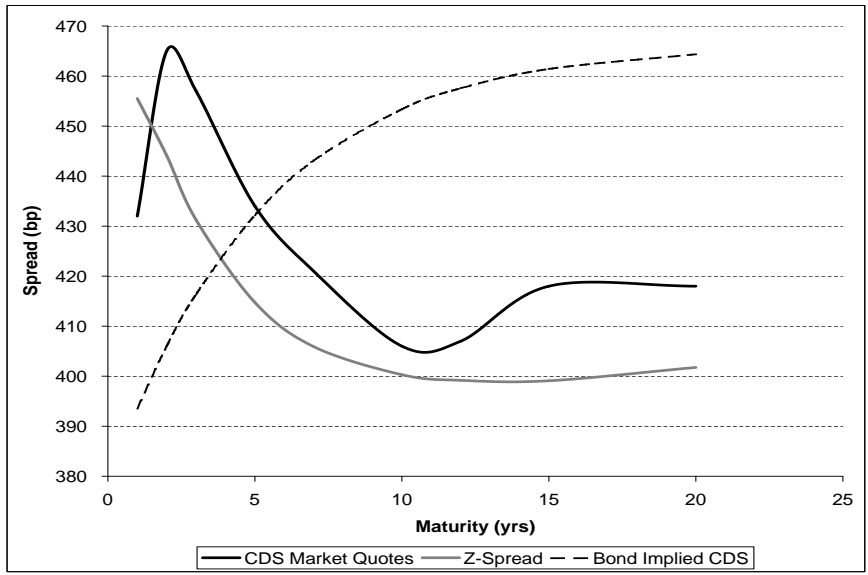}%
\caption{{\small Z-spread, CDS and BCDS term structures. Ford Motor Credit, 12/31/2002.}}%
\label{fig:optical-inversion-cds}%
\end{figure}

Figure \ref{fig:optical-inversion-cds} shows the Z-spread, Credit Default Swap (CDS), and the bond-implied CDS (BCDS, defined in section \ref{sec:BCDS} below) term structures of Ford for the same date. From the figure it is clear that investors who looked at the Z-spread term structure and compared it to the cost of protection available via the credit default swap (CDS) market, would have been misled to think that there was a small positive basis between the two, i.e. that the CDS traded wider by roughly 20 bp than bonds across most maturities, with the shape of the curve following closely the shape of the Z-spread curve with the exception of less than 2 year maturities. In fact, if one had used the methodology presented in this chapter and derived the BCDS term structure and compared it with the market quotes, one would see a very different picture - the bond market traded more than 50 bp tighter than CDS at short maturities and more than 50 bp wider at maturities greater than 5 years, if measured on an apples-to-apples basis. 

Both the bonds and the CDS of Ford Motor Credit are among the most liquid and widely traded instruments in the U.S. credit market and these differences are clearly not a reflection of market inefficiency but rather of the difference between the methodologies. Only for the very short term exposures, where investors are forced to focus critically on the likelihood of the default scenarios and the fair value of the protection with a full account for cash flow outcomes do we see the CDS market diverging in shape from the cash market's optically distorted perception of the credit spreads. 

\section{The Survival-Based Valuation of Credit Risky Securities}

In this section we outline the credit valuation methodology which follows an assumption, known as the {\em reduced form framework}, that while the default risk is measurable and anticipated by the market, the timing of the actual event of the default is exogenous and unpredictable. This assumption critically differs from a more fundamental approach taken in {\em structural models} of credit risk \cite{Merton-1974}, and where not only the level of default risk but also its timing become gradually more predictable as the company nears the default barrier. Within this framework, we argue that only the so-called Fractional Recovery of Par (FRP) assumption is consistent with the market practices. We call this particular version of the reduced form framework combined with the FRP assumption the {\em survival-based valuation framework}.

\subsection{The Single-Name Credit Market Instruments} \label{sec:single_name_credit}

We will focus in our exposition on the single-name credit market (the multi-name structured credit market is discussed elsewhere in this book). The basic instruments of this market are:

\begin{description}

\item[Credit Bond (CB)] is a a security, which entitles the investor to receive regular (typically semi-annual) coupon payments plus a return of the principal at the end of the maturity period. The security is purchased by making a cash payment upfront. Typically, the security is issued by a corporation or another entity borrowing the money, and represents a senior (compared to equity) claim against the assets of the borrower. 

\item[Credit Default Swap (CDS)] is a contract between two counterparties, where the protection buyer makes regular (typically quarterly) premium payments until the earlier of the maturity or credit event, plus possibly a single upfront payment at the time of entering in the contract. In exchange for these, he expects to receive from the protection seller a single payment in case of credit event (default, bankruptcy or restructuring) prior to maturity, which is economically equivalent to a making up the difference between the par value of the referenced credit bond(s) and their post-default market value, known as the recovery price\footnote{Technical details include the definition of the deliverable basket of securities under the ISDA terms, netting of accrued payments, and the mechanism of settling the protection payment which can be via physical delivery or auction-driven cash settlement \cite{OKane-book}.}. 

\item[Digital Default Swap (DDS)] is similar to the cash-settled CDS in most respects, except that the protection payment amount is contractually predefined and cited in terms of contractual recovery rate. For example, if the contractual recovery rate is set to zero, the protection payment would be equal to the contract notional. 

\item[Constant Maturity Default Swap (CMDS)] is similar to conventional CDS, except that the periodic premium payment amount is not fixed but rather floating and is linked to some benchmark reference rate, such as the 5-year par CDS rate of the same or related entity. CMDS reduces the mark-to-market risk of the CDS with respect to credit spread fluctuations in the same way as the floating rate notes reduce the mark-to-market risk of bullet bonds with respect to interest rate fluctuations.  

\item[Recovery Swap (RS)] is a contract between two counterparties whereby they agree to exchange the realized recovery vs. the contractual recovery value (recovery swap rate) in case of default, with no other payments being made in any other scenario. The typical recovery swaps have no running or upfront payments of any kind. In keeping with the swaps nomenclature, the counterparties are denoted as the `payer' and `receiver' of the realized recovery rate, correspondingly \cite{Berd-RecoverySwaps}. 

\end{description}

The conventional credit bonds and CDS still remain the linchpin of the credit market and continue to account for the majority of all traded volumes, according to the most recent statistics from the Bank of International Settlement \cite{BIS-2009}. But the existence of the expanded toolkit including the (niche markets for) CMDS, DDS and RS implies certain connections and (partial) substitution ability between various credit derivatives instruments, which we will study in section \ref{sec:credit_triangle}. 

The relationship between credit bonds and CDS is also non-trivial and leads to existence of a pricing basis between the two liquid markets. While representing opportunities for relative value investment and even occasionally a true arbitrage, this basis remains driven by market segmentation and strong technical factors. We will discuss the CDS-Bond basis in detail and provide tools to analyze it in section \ref{sec:cds_bond_basis}.

\subsection{The Recovery Assumptions in the Reduced-Form Framework} \label{sec:recovery_assumptions}

The reduced-form framework for valuation of credit-risky bonds had a long history of development -- see the pioneering works by Litterman and Iben \cite{Litterman-Iben-1991}, Jarrow and Turnbull \cite{Jarrow-Turnbull-1995}, Jarrow, Lando and Turnbull \cite{Jarrow-Lando-Turnbull-1997} and Duffie and Singleton \cite{Duffie-Singleton-1999} as well as the textbooks by Duffie and Singleton \cite{Duffie-Singleton-book}, Lando \cite{Lando-book}, Schonbucher \cite{Schonbucher-book}, and O'Kane \cite{OKane-book} for detailed discussions and many more references. The key assumption in these models is what will be the market value of the bond just after the default. In other words, one must make an assumption on what is the expected recovery given default (or alternatively what is the loss given default). There are three main conventions regarding this assumption: 

\begin{description}
\item[Fractional recovery of Treasury (FRT):] In this approach, following \cite{Jarrow-Turnbull-1995}, one assumes that upon default a bond is valued at a given fraction to the hypothetical present value of its remaining cash flows, discounted at the riskless rate.
\item[Fractional recovery of market value (FRMV):] Following \cite{Duffie-Singleton-1999}, one assumes in this approach that upon default a bond loses a given fraction of its value just prior to default.
\item[Fractional recovery of par (FRP):] Under this assumption, a bond recovers a given fraction of its face value upon default, regardless of the remaining cash flows. A possible extension is that the bond additionally recovers a (possibly different) fraction of the current accrued interest.
\end{description}

Both FRMV and FRT assumptions lead to very convenient closed form solutions for pricing defaultable bonds as well as derivatives whose underlying securities are defaultable bonds. In both approaches one can find an equivalent `risky' yield curve which can be used for discounting the promised cash flows of defaultable bonds and proceed with valuation in essentially the same fashion as if the bond was riskless - the only difference is the change in the discounting function. As a result, either of these approaches works quite well for credit-risky bonds that trade not too far from their par values (see \cite{Jarrow-Turnbull-2000} for a related discussions). 

The main drawback of both of these approaches is that they do not correspond well to the market behaviour when bonds trade at substantial price discounts. Namely, the FRMV assumption fails to recognize the fact that the market begins to discount the future recovery when the bonds are very risky. In other words, the bonds are already {\em trading to recovery} just prior to default, therefore there is often relatively little additional market value loss when the actual event of default takes place. 

The FRT assumption, on the other hand, does not recognize the acceleration of debt and we would argue is wholly unsuitable for the valuation of credit risky instruments. 

Of course, both FRMV and FRT approaches can be adjusted to conform to market behaviour by generalizing the expected recovery from a constant to a variable dependent on the current price of the bonds. However, such a generalization would invalidate the closed-form expressions for risky yields and negate the main advantage of these recovery assumptions. In fact, we think that what is normally considered to be an advantage of the FRMV and FRT recovery assumptions is actually a deficiency. Namely, the possibility of a strippable cash flow valuation under these assumptions with the present value of a credit bond being a simple sum of the present values of contractual cash flows is in contradiction with our understanding that all credit bonds, regardless of their contractual structure, have an embedded option to default and therefore they simply cannot be thought of as just a portfolio of coupon and principal cash flows - irrespective of whether the inherent option to default is exercised rationally by a limited liability issuer or is triggered by exogenous factors. 

The only limiting case when the FRMV and FRT assumptions are in agreement with the acceleration of debt and equal priority recovery is when the expected recovery value is precisely equal to zero. We denote this as the zero recovery (ZR) assumption. As we will see in the subsequent sections, these three cases (FRMV, FRT and ZR recovery assumptions) are the only cases when the valuation of a bond as a portfolio of individual contractual cash flows remains valid despite the possibility of a default scenario. This is simply due to the fact that under these assumptions one does not introduce any new cash flow values which were not already present in bond's contractual description. Therefore, when calculating the present value of a credit bond, one can combine the riskless discount function, the survival probability, and the assumed recovery fraction into a {\em risky discount function} which can then be applied to the contractual cash flows (see \cite{Duffie-Schroeder-Skiadas-1996} for a discussion of conditions under which the valuation of defaultable securities can be performed by applying a risky stochastic discount process to their default-free payoff stream).

In contrast, the fractional recovery of par (FRP) assumption is fully consistent with the market dynamics, and can explain some of the salient features of the distressed credit pricing in a very intuitive manner as discussed in the previous section. During the recessions of 2001-2002 and 2008-2009 a large number of fallen angel issuer bonds of various maturities have been trading at deep discounts. The analysis of this large set of empirical data, whose results are partially reported in this review, confirms that the FRP recovery assumption indeed leads to a more robust modeling framework compared with the FRMV and FRT assumptions. 

Despite the widespread use of alternative recovery assumptions by practitioners and academics, there are only a handful of studies which examine their importance for pricing of standard credit bonds and CDS. Finkelstein \cite{Finkelstein-1999} has pointed out the importance of the correct choice of the recovery assumption for the estimation of the term structure of default probabilities when fitted to observed CDS spreads, and the fact that the strippable valuation of credit bonds becomes impossible under the FRP assumption. 

Duffie \cite{Duffie-1998} has explored the pricing of default-risky securities with fractional recovery of par. In particular, he derived a generic result relating the risk-neutral implied hazard rate to the short spread via the widely used credit triangle formula  $h=S/(1-R)$. The important question, however, is what is the meaning of the spread used in this relationship. Under the assumptions in \cite{Duffie-1998}, this is the spread of a zero-coupon credit-risky principal strip - an asset that does not actually exist in marketplace. In contrast, we develop a FRP-based pricing methodology with alternative definition of spreads that refers to either full coupon-bearing bonds or CDS.

Bakshi, Madan and Zhang \cite{Bakshi-Madan-Zhang-2006} have specifically focused on the implications of the recovery assumptions for pricing of credit risk. Having developed a valuation framework for defaultable debt pricing under all three recovery assumptions (FRMV, FRT and FRP), they have concluded from the comparison with time series of 25 liquid BBB-rated bullet bonds that the FRT assumption fits the bond prices best. While we agree with their methodology in general, we believe that in this case that the {\em market is wrong} for a variety of legacy reasons discussed in the previous section, and one must insist on a better FRP model {\em despite} the empirical evidence from the investment grade bond prices. Our own estimates, based on 15 years of monthly prices for more than 5000 senior unsecured bonds across all rating categories, available for review via Barclays Capital Quantitative Credit Toolkit, suggest that the FRP assumption allows for a more robust fit across a larger sample.

Das and Hannona \cite{Das-Hannona} have extended the pricing framework by introducing a non-trivial implied recovery rate process. They have derived pricing formulae for credit bonds under various assumptions of the recovery rate dynamics. Their extended framework allows to incorporate negative correlations of the recovery rate with the default process, achieving a closer agreement with the empirical observations \cite{Altman-2005}. In our review, we will retain a simpler constant recovery rate assumption in order to derive intuitive relative value measures, most of which refer to a static set of bond and CDS prices at a given time and do not concern the dynamics of recovery rates.

Finally, in an important empirical study Guha \cite{Guha-2003} examined the realized recoveries of U.S.-based issuers and concluded that the FRP assumption is strongly favored by the data in comparison to the FRMV or FRT assumptions. In particular, he has shown that the vast majority of defaulted bonds of the same issuer and seniority are valued equally or within one dollar, irrespective of their remaining time to maturity.

\subsection{Pricing of Credit Bonds} \label{sec:pricing_FRP}

Consider a credit-risky bond with maturity $T$ that pays fixed cash flows with specified frequency (usually annual or semi-annual). According to the fractional recovery of par assumption, the present value of such a bond is given by the expected discounted future cash flows, including the scenarios when it defaults and recovers a fraction of the face value and possibly of the accrued interest, discounted at the risk-free (base) rates. By writing explicitly the scenarios of survival and default, we obtain the following pricing relationship at time t:

\begin{eqnarray} \label{eq:pricing-FRP}
	PV(t) & = & \sum_{t_{i}>t}^{t_{N}}{ \left( CF_{pr}(t_{i}) + CF_{int}(t_{i}) \right) \, E_{t}\left\{ Z_{t_{i}} \, 1_{t_{i} < \tau} \right\} } \nonumber \\
	      & + & \int_{t}^{T} {E_{t} \left\{ R_{pr} \, F_{pr}(\tau) \, Z_{\tau} \, 1_{u< \tau \leq u+du} \right\}} \nonumber \\ 
	      & + & \int_{t}^{T} {E_{t} \left\{ R_{int} \, A_{int}(\tau) \, Z_{\tau} \, 1_{u< \tau \leq u+du} \right\}} 
\end{eqnarray}

The variable $\tau$ denotes the (random) default time, $1_{X}$ denotes an indicator function for a random event X, $Z_{t}$ is the (random) base discount factor, and   $E_{t}\left\{\cdot \right\}$ denotes the expectation under the risk-neutral measure at time $t$.   

The first sum corresponds to scenarios in which the bond survives until the corresponding payment dates without default. The total cash flow at each date is defined as the sum of principal  $CF_{pr}$, and interest  $CF_{int}$, payments. The integral corresponds to the recovery cashflows that result from a default event occurring in a small time interval  $u< \tau \leq u+du$ , with the bond recovering a fraction  $R_{pr}$ of the outstanding (amortized) principal face value $F_{pr}(\tau)$  plus a (possibly different) fraction $R_{int}$  of the interest accrued  $A_{int}(\tau)$.  

Assuming the independence of default times, recovery rates and interest rates\footnote{For alternative assumptions see Das and Hannona \cite{Das-Hannona}, who considered the problem of credit pricing under correlated default and recovery rates, and Jarrow and collaborators \cite{Jarrow-2001, Jarrow-Yildirim-2002} who considered the correlated interest and hazard rates. Our simplified assumption of independence remains, however, among the more popular conventions used both in academia and among the practitioners.}, one can express the risk-neutral expectations in eq. (\ref{eq:pricing-FRP}) as products of separate factors encoding the term structures of (non-random) base discount function, survival probability and conditional default probability, respectively:

\begin{eqnarray} \label{eq:pricing-functions}
Z_{base}(t,u) & = & E_{t}\left\{ Z_{u} \right\} \label{eq:pricing-functions-1} \\
Q(t,u)  & = & E_{t}\left\{ 1_{u < \tau} \right\} \label{eq:pricing-functions-2} \\
D(t,u)  & = & \lim_{du \rightarrow 0} \frac{E_{t}\left\{ 1_{u< \tau \leq u+du} \right\}}{du} = - \frac{d}{du} Q(t,u) \label{eq:pricing-functions-3}
\end{eqnarray}

Many practitioners simply use a version of the equation (\ref{eq:pricing-FRP}) assuming that the recovery cash flows occurs on the next coupon day $\tau = t_{i}$, given a default at any time within the previous coupon payment period $[t_{i-1},t_{i}]$. As a possible support for this assumption, one might argue that the inability to meet company's obligations is more likely to be revealed on a payment date than at any time prior to that when no payments are due, regardless of when the insolvency becomes inevitable. Of course, this argument becomes much less effective if the company has other obligations besides the bond under consideration. Still, to simplify the implementation we will follow this approach in the empirical section of this chapter. 

We assume that the unpaid accrued interest is added to the outstanding principal in case of default, as is the common practice in the US bankruptcy proceedings, and set $R_{int}=R_{pr}=R$. Any potential inaccuracy caused by these assumptions is subsumed by the large uncertainty about the level of the principal recovery, which is much more important.

For the case of fixed-coupon bullet bonds with coupon frequency $q$ (e.g. semi-annual $q=2$), the average timing of default is half-way through the coupon period, and the expected accrued interest amount is half of the next coupon payment. This gives a simplified version of the pricing equation where we suppressed the time variable $t$ (see also Appendix \ref{sec:continuous_time} for the continuous time approximation):

\begin{eqnarray} \label{eq:pricing-FRP-simple}
	PV & = & Z_{base}(t_{N}) \, Q(t_{N}) + \frac{C}{q} \sum_{i=1}^{N}{ Z_{base}(t_{i}) \, Q(t_{i}) } \nonumber \\
	      & + & R \, \left(1 + \frac{C}{2q} \right) \, \sum_{i}^{N}{ Z_{base}(t_{i}) \, \left( Q(t_{i-1}) - Q(t_{i}) \right) }
\end{eqnarray}

Here, the probability $D(t_{i-1},t_{i})$  that the default will occur within the time interval $[t_{i-1},t_{i}]$, conditional on surviving until the beginning of this interval, is expressed through the survival probability in a simple way:

\begin{equation} \label{eq:default-prob-finite}
D(t_{i-1},t_{i}) = Q(t_{i-1}) - Q(t_{i})
\end{equation}

One can see quite clearly from this expression that under the fractional recovery of par (FRP) assumption the present value of the coupon-bearing credit bond does not reduce to a simple sum of contractual cash flow present values using any risky discount function. An obvious exception to this is the case of zero recovery assumption, under which the survival probability plays the role of a risky discount function.

\subsection{Pricing of CDS} \label{sec:pricing_CDS}

Unlike credit bonds, CDS have always been priced with FRP assumption, for a simple reason that the default scenario is central in the definition of this instrument. The credit default swap consists of two legs, premium leg which corresponds to regular (typically quarterly) payments of the contractual premium amount by the buyer of protection, and the protection leg which corresponds to the contingent payment of the default payoff amount to the protection buyer in case of a qualified credit event. 

The most generic case of a CDS which trades with an upfront payment $UP$ amount and contractual premium payments $CF_{prem} = C_{CDS}/q$ is priced by requiring that the present value of the premium leg be equal to the present value of the protection leg, including the market standard convention for netting of the accrued premium $A_{prem}$ and the principal protection payment $1-R_{pr}$:

\begin{eqnarray} \label{eq:pricing-CDS}
	UP(t) & + & \frac{C_{CDS}}{q} \, \sum_{t_{i}>t}^{t_{N}}{ E_{t} \left\{ Z_{t_{i}} \, 1_{t_{i} < \tau} \right\} } \nonumber \\
	      & = &  \int_{t}^{T} {E_{t} \left\{ \left(1 - R_{pr} - A_{prem,\tau} \right) \, Z_{\tau} \, 1_{u< \tau \leq u+du} \right\} } 
\end{eqnarray}

Making a similar set of simplifications as in the case of credit bonds, we get (see also Appendix \ref{sec:continuous_time} for the continuous time approximation):

\begin{eqnarray} \label{eq:pricing-CDS-simple}
	 UP & + & \frac{C_{CDS}}{q} \, \sum_{i=1}^{N}{ Z({t_{i}}) \, Q({t_{i}}) } \nonumber \\
	 		& = & \left(1 - R - \frac{C_{CDS}}{2q} \right) \, \sum_{i=1}^{N}{ Z({t_{i}}) \, \left( Q({t_{i-1}}) - Q({t_{i}}) \right) }
\end{eqnarray}

If the upfront payment is equal to zero, the CDS is said to be trading at par, and its coupon then coincides with the par CDS spread, defined as follows:

\begin{equation} \label{eq:pricing-CDS-spread}
	 S_{CDS} =
	   2 q \, \left(1 - R\right) \, \frac{\sum_{i=1}^{N}{ Z({t_{i}}) \, \left( Q({t_{i-1}}) - Q({t_{i}}) \right) }}{\sum_{i=1}^{N}{ Z({t_{i}}) \, \left( Q({t_{i-1}}) + Q({t_{i}}) \right) }}
\end{equation}

The equation (\ref{eq:pricing-CDS-spread}) defines the par (or breakeven) spread for CDS even if its contractual coupon is such that the upfront payment is non-zero. It is the best measure of relative value for comparing different CDS of the same maturity. One can express the amount of the upfront payment (which could also be interpreted as the mark-to-market value of CDS) through the difference between the par CDS spread $S_{CDS}$ and the contractual premium $C_{CDS}$:

\begin{eqnarray} 
	UP    & = & \left(S_{CDS} - C_{CDS} \right) \, \pi \label{eq:mtm-CDS} \\
	\pi   & = &  \frac{1}{2q} \, \sum_{i=1}^{N}{ Z({t_{i}}) \, \left(Q({t_{i-1}}) + Q({t_{i}}) \right) } \label{eq:rpv01-CDS}
\end{eqnarray}
where $\pi$ is known as the `risky PV01', or the risky price of a basis point.

After the recent modifications of the CDS contract conventions, all single-name CDS trade with fixed coupons of either 100 bps or 500 bps plus appropriate upfront payment. As seen from (\ref{eq:mtm-CDS}), in case if the par spread is less than the contractual coupon, the upfront payment will actually be negative, i.e.\ the counterparty purchasing the protection will {\em receive} an upfront payment which will compensate it for greater-than necessary premium payments in the future.

The critical question in these equations is, of course, the estimates of the survival probability $Q(t)$. Next section shows that it is directly related to CDS spread and recovery value via a so-called `credit triangle' formula. 

\subsection{The Credit Triangle and Default Rate Calibration} \label{sec:credit_triangle}

The phrase {\em credit triangle} refers to the relationship between the credit spread, default rate and recovery rate. It is often cited as follows:

\begin{equation} \label{eq:TriangleTraditional}
	\mbox{Credit Spread} = \mbox{Default Probability Rate} \times \left( 1 - \mbox{Recovery Rate} \right)
\end{equation}

The appealing simplicity of this formula masks some ambiguities associated with its interpretation. Within the reduced form model, the default probability rate is the hazard rate of the exogenous Poisson process of default event arrival, the recovery rate is a model parameter which can be considered constant or stochastic, and the credit spread is the combined measure of credit risk that remains open to interpretation depending on which particular assumption of recovery is assumed or which particular credit instrument is considered. Outside of the model framework, the practitioners often (mis)use this formula by plugging into it the yield spread or the Z-spread of credit bonds, which are inconsistent measures of credit risk as explained in this chapter. 

Rather than defining these metrics within a particular modeling framework, we propose to specify them in terms of observable quantities of tradable instruments such as credit default swaps (CDS), digital default swaps (DDS) and recovery swaps (RS). The no-arbitrage relationship between the contractual rates of these instruments is the unambiguous alternative to the formula (\ref{eq:TriangleTraditional}).

For example, the recovery swaps can be fully replicated by a combination of conventional and digital CDS. This replication, as usual, implies an arbitrage-free relationship between these three instruments which we will derive below. 

Consider the replication trade depicted in Table \ref{tab:StaticReplication}: a unit notional payer recovery swap with the fixed swap rate $R_{RS}$ is hedged with a long protection position in DDS with notional $H_{DDS}$ and short protection in CDS with a notional $-H_{CDS}$. The cash flows columns indicate the cash flows per unit notional on each leg of the trade. 

\begin{table}[tp]
\begin{tabular}{|r|c|c|c|c|c|} \hline
					& Notional 	& \multicolumn{4}{|c|}{Cash Flows} 	\\ \hline
					& 				 	& Upfront & Premium & Default 				& Maturity 	\\ \hline \hline
RS				& $+1$				& $0$			& $0$			&	$R_{RS} - R$		&	$0$								\\ \hline
DDS 			& $+H_{DDS}$	& $0$			& $-S_{DDS}$	&	$1 - R_{DDS}$	&	$0$								\\ \hline
CDS 			& $-H_{CDS}$	& $0$			& $-S_{CDS}$	&	$1 - R$				&	$0$								\\ \hline \hline
Net				& 					& $0$			& $ \begin{array}{c} -H_{DDS} \, S_{DDS} \\ + H_{CDS} \, S_{CDS} \end{array}$	&	$\begin{array}{c} R_{RS} - R \\ + H_{DDS} \, (1 - R_{DDS}) \\ - H_{CDS} \, (1 - R) \end{array}$				&	$0$								\\ \hline
\end{tabular}
	\caption{Static Replication of Recovery Swaps by CDS and DDS}
	\label{tab:StaticReplication}
\end{table}

The net cash flows are identically zero for upfront payments. The hedge ratios $H_{DDS}$ for the DDS and $H_{CDS}$ for the CDS should be chosen so that the net cash flows are also zero for premium and default payments. Consider the case of default:

\begin{equation} \label{eq:rs_cf_default}
CF_{default} = R_{RS} - R + H_{DDS} \, (1 - R_{DDS}) - H_{CDS} \, (1 - R)
\end{equation}

In order to guarantee that these cash flows are equal to zero regardless of the realized recovery rate $R$, one must have:

\begin{eqnarray} \label{eq:HedgeRatios}
H_{CDS} & = & 1 \\
H_{DDS} & = & \frac{1-R_{RS}}{1-R_{DDS}} 
\end{eqnarray}

Consider now the net premium cash flows:

\begin{equation} \label{eq:rs_cf_premium}
CF_{prem} = -H_{DDS} \, S_{DDS} + H_{CDS} \, S_{CDS}
\end{equation}

Because all other cash flows of the replication trade are identically zero, then plugging the hedge ratios (\ref{eq:HedgeRatios}) into (\ref{eq:rs_cf_premium}) and requiring that the net premium cash flows be also equal to zero results in a no-arbitrage relationship between the matching maturity $T$ recovery swaps, CDS and DDS rates: 

\begin{equation} \label{eq:NoArbTriangle}
S_{DDS}(T) = \frac{1-R_{DDS}(T)}{1-R_{RS}(T)} \, S_{CDS}(T)
\end{equation}

Based on this relationship, it would be natural to use the recovery swap rates, rather than historical estimates or other forecasts of the realized recovery rate, for calibration of the implied hazard rates within the risk-neutral framework. 

Indeed, imagine that we have observed the term structure of CDS spreads and recovery swap rates for a range of maturities $[0, T]$. According to (\ref{eq:NoArbTriangle}) we can obtain without any further assumptions the term structure of DDS spreads with zero contractual recovery $R_{DDS}=0$. The pricing of such DDS depends solely on the term structure of risk-neutral discount rates and survival probabilities and possibly on the correlation of the default process and the risk-free discount factor (compare with (\ref{eq:pricing-CDS})):

\begin{equation} \label{eq:pricing-DDS}
S_{DDS}(T) \, \int_{0}^{T}{du \, E\left\{Z_{u} \, 1_{u < \tau}\right\}} = \int_{0}^{T}{E\left\{Z_{u} \, 1_{u < \tau \leq u+du}\right\}}
\end{equation}

For convenience in notations, we have adopted an approximation of continuous DDS premium payments (this approximation is irrelevant for the subsequent discussion). Under the assumption of independence between the exogenous default process and the risk-free rates, we get:

\begin{equation} \label{eq:CalibrateDDS}
\frac{S_{CDS}(T)}{1-R_{RS}(T)} \, \int_{0}^{T}{du \, Q(u) \, Z(u)} = \int_{0}^{T}{du\, h(u) \, Q(u) \, Z(u)}
\end{equation}

One can easily calibrate the term structure of implied hazard rates $h(s)$ to the market-observed CDS spreads and recovery swap rates from (\ref{eq:CalibrateDDS}) without making any additional assumptions about the recovery and default process. For example, in the case of flat CDS spreads, recovery swap rates and hazard rates one immediately obtains the conventional `credit triangle' (compare with (\ref{eq:TriangleTraditional})):

\begin{equation} \label{eq:CreditTriangleDerived}
h = \frac{S_{CDS}}{1-R_{RS}}
\end{equation}

This confirmes that the no-arbitrage relationship (\ref{eq:NoArbTriangle}) is the unambiguous market-based generalization of the familiar credit triangle. 

Note that in our derivation of the equations (\ref{eq:CalibrateDDS}) and (\ref{eq:CreditTriangleDerived}) we did not make an assumption about the existence of a liquid DDS market -- it is sufficient to assume that the CDS and recovery swaps markets exist, and price the hypothetical DDS based on the known arbitrage-free relationship between the DDS and CDS spreads (\ref{eq:NoArbTriangle}). Alternatively, if we assume the existence of CDS and DDS market, the market triangle equation (\ref{eq:NoArbTriangle}) directly defines the correct recovery swap rate, which should also be used in other instances wherever a risk-neutral `implied recovery rate' is needed. In any case, only two out of three elements of the market triangle containing CDS, DDS and RS need be observable.

Unfortunately, the state of the market is such that RS or DDS contracts are rarely traded, often only for distressed issuers where the default scenario is very likely. This leaves a lot of ambiguity for the choice of the recovery rate to be used in the calibration of the implied hazard rates. The credit triangle formula states that the calibrated implied hazard rate is a growing function of assumed recovery rate, given a fixed level of observed CDS spreads $S_{CDS}$. 

This is in a stark contrast with the empirical evidence of a negative correlation between historical default rates and recovery rates \cite{Altman-2005}. Such difference in behavior should not be puzzling since we are comparing the static dependence of implied hazard rates on implied recovery rates conditioned on constant CDS spread with dynamic historical rates that are not so conditioned. In fact, the calibration procedure for the implied hazard rates using the recovery swap rates is completely independent of assumptions regarding the correlation between the recovery rates and default events, since the realized recovery rate is simply absent from equation (\ref{eq:pricing-DDS}). The readers can refer to \cite{Bakshi-Madan-Zhang-2006} and \cite{Das-Hannona} for an expanded modeling framework including variable hazard rates and recovery rates, where the correlation between them is a tunable parameter and thus can match the observed negative correlation.

\section{Empirical Estimation of Survival Probabilities}

Having derived the pricing relationship in the survival-based approach, we are now ready to estimate the implied survival probability term structure directly from the bond prices. The premise of our approach is that the survival probability is an exponentially decaying function of maturity, perhaps with a varying decay rate. This assumption is generally valid in Poisson models of exogenous default, where the default hazard rate is known but the exact timing of the default event is unpredictable. This is the same assumption made by all versions of reduced-form models regardless of the recovery assumptions discussed earlier. Notably, this assumption differs from the Merton-style structural models of credit risk \cite{Merton-1974} where the timing of default becomes gradually more predictable as the assets of the firm fall towards the default threshold. 

When it comes to the estimation of term structures based on a large number of off-the-run bonds across a wide range of maturities, most approaches based on yield or spread fitting are not adequate because they lead to a non-linear dependence of the objective function on the fit parameters. The most important aspect of this problem is the large number of securities to be fitted which makes a precise fit of all prices impractical (or not robust) and therefore creates a need for a clear estimate of the accuracy of the fit. After all one must know whether a given bond trading above or below the fitted curve represents a genuine rich/cheap signal or whether this mismatch is within the model's error range. Without such estimate relative value trading based on fitted curves would not be possible.

Vasicek and Fong \cite{Vasicek-Fong-1982} (see also \cite{Shea-1985}) suggested a solution to this dilemma, which has become a de-facto industry standard for off-the-run Treasury and agency curve estimation . They noted that the above problem is best interpreted as a cross-sectional regression. As such, it would be best if the explanatory factors in this regression were linearly related to the observable prices, because this would lead to a (generalized) linear regression. Realizing further that the quantity which is linearly related to bond prices is the discount function, they proposed to estimate the term structure of risk-free discount function itself rather than the term structure of yields. Finally, they argued that the simplest discount function is exponentially decreasing with a constant rate, and concluded that one must use exponential splines, which are linear combination of exponential functions, to best approximate the shape of realistic discount functions. We review the definition of exponential splines in the Appendix \ref{sec:splines}.

In the case of credit risky bonds a similar logic also applies, except that one has to think about the survival probabilities rather than discount function, because it is the survival probabilities that appear linearly in the bond pricing equation (\ref{eq:pricing-FRP-simple}). When the hazard rate is constant the survival probability term structure is exactly exponential. Therefore, it is indeed well suited for approximation by exponential splines (ref{eq:exp-spline}):

\begin{equation} \label{eq:survival-spline}
Q(t) = \sum_{k=1}^{K}{\beta_{k} \, \Phi_{k}\left(t\left|\eta \right. \right)}
\end{equation}
where the spline factors $\Phi_{k}\left(t\left|\eta\right. \right)$  depend on the tenor $t$  and on the long-term decay factor $\eta$, which in this case has the meaning of the generic long-term default hazard rate $h(t \rightarrow \infty) = \eta$. 

Assuming that we have already estimated the base discount function, and substituting the spline equation (\ref{eq:survival-spline}) into pricing equation (\ref{eq:pricing-FRP-simple}) we obtain a cross-sectional regression setting for direct estimation of the survival probability term structure from observable bond prices: 

\begin{equation} \label{eq:survival-regression}
V_{j} = \sum_{k=1}^{K}{\beta_{k} \, U_{j,k} + \epsilon_{n}}
\end{equation}
which, assuming $k=3$, in matrix notations looks like:

\begin{equation} \label{eq:survival-regression-matrix}
\left[\begin{array}{c} V_{1} \\ \vdots \\ V_{n} \\ \vdots \end{array} \right] = 
\left[\begin{array}{ccc} U_{1,1} & U_{1,2} & U_{1,3} \\
\vdots & \vdots & \vdots \\
U_{j,1} & U_{j,2} & U_{j,3} \\
\vdots & \vdots & \vdots 
\end{array} \right] 
\cdot 
\left[\begin{array}{c} \beta_{1} \\ \beta_{2} \\ \beta_{3} \end{array} \right]
 + 
\left[\begin{array}{c} \epsilon_{1} \\ \vdots \\ \epsilon_{j} \\ \vdots \end{array} \right] 
\end{equation}

Here we introduced the explanatory variables $U_{j,k}$ for the $j$-th bond and $k$-th spline factor and the adjusted present value $V_{j}$ for the $j$-th bond as:

\begin{eqnarray} \label{eq:survival-regression-rhs}
	U_{j,k} & = & \sum_{i=1}^{N}{ \Phi_{k}\left(t_{j,i}\left|\eta \right. \right) \, \left( \frac{C_{j}}{q_{j}} Z(t_{j,i}) - R  \, \left(1 + \frac{C_{j}}{2q_{j}} \right) \, \left(Z(t_{j,i}) - Z(t_{j,i+1}) \right) \right) } \nonumber \\
	      & + & Z(t_{j,N}) \, \Phi_{k}\left(t_{j,N}\left|\eta \right. \right) \, \left( \frac{C_{j}}{q_{j}} + 1 - R  \, \left(1 + \frac{C_{j}}{2q_{j}} \right) \right) 
\end{eqnarray}
	 
\begin{equation} \label{eq:survival-regression-lhs}
V_{j} = PV_{j} - R \, \left(1 + \frac{C_{j}}{2q_{j}} \right) \, Z(t_{j,1})
\end{equation}

We have found empirically that it is often sufficient to retain only the first three factors for estimating the survival probability. Thus there are no knot-factors in our implementation of this approach. The first three coefficients of the spline expansion must satisfy an equality constraint, because the survival probability must be exactly equal to 1 when the time horizon is equal to zero.

\begin{equation} \label{eq:survival-regression-constraint1}
\sum_{k=1}^{3}{ \beta_{k} } = 1
\end{equation}

In addition to the equality constraint, we also impose inequality constraints at intermediate maturities $T_{c}$  to make sure that the survival probability is strictly decreasing, and consequently the hazard rate is strictly positive. Their functional form is:

\begin{equation} \label{eq:survival-regression-constraint2}
\sum_{k=1}^{3}{ \beta_{k} \, k \, e^{-k \eta T_{c}}} > 0
\end{equation}

In addition, we impose a single constraint at the long end of the curve to make sure that the survival probability itself is positive. 

\begin{equation} \label{eq:survival-regression-constraint3}
\sum_{k=1}^{3}{ \beta_{k} \, e^{-k \eta T^{max}_{c}}} > 0
\end{equation}

Together with the strictly decreasing shape of the survival probability term structure guaranteed by (\ref{eq:survival-regression-constraint2}), this eliminates any possibility of inconsistency of default and survival probabilities in the exponential spline approximation:

It is worth noting that in most cases the inequality constraints (\ref{eq:survival-regression-constraint2}), (\ref{eq:survival-regression-constraint3}) will not be binding and therefore the regression estimates will coincide with the simple GLS formulas. The constraints will kick in precisely in those cases where the input data is not consistent with survival-based modeling, which can happen for variety of reasons including the imperfection of market pricing data, company-specific deviations of expected recovery rates, etc.

We use a two-tiered weighting scheme for the regression objective function with the first set of weights inversely proportional to the square of the bond's spread duration $SD_{j}$ to make sure that the relative accuracy of the hazard rate estimates is roughly constant across maturities. The second set of weights is iteratively adjusted to reduce the influence of the outliers following the generalized M-estimator technique described in Wilcox \cite{Wilcox-book}.

\begin{equation} \label{eq:survival-regression-OF}
OF = \sum_{j=1}^{J_{bonds}}{ \frac{w^{outlier}_{j}}{\sqrt{SD_{j}}} \, \epsilon_{j}^{2}}
\end{equation}

Equations (\ref{eq:survival-spline}) -- (\ref{eq:survival-regression-OF}) fully specify the estimation procedure for survival probability term structure. It satisfies the main goals that we have defined at the outset - the procedure is robust, it is consistent with market practices and reflects the behavior of distressed bonds, and is guaranteed to provide positive default probabilities and hazard rates.

\begin{figure}[t]
\includegraphics[height=3in,width=4.5in]{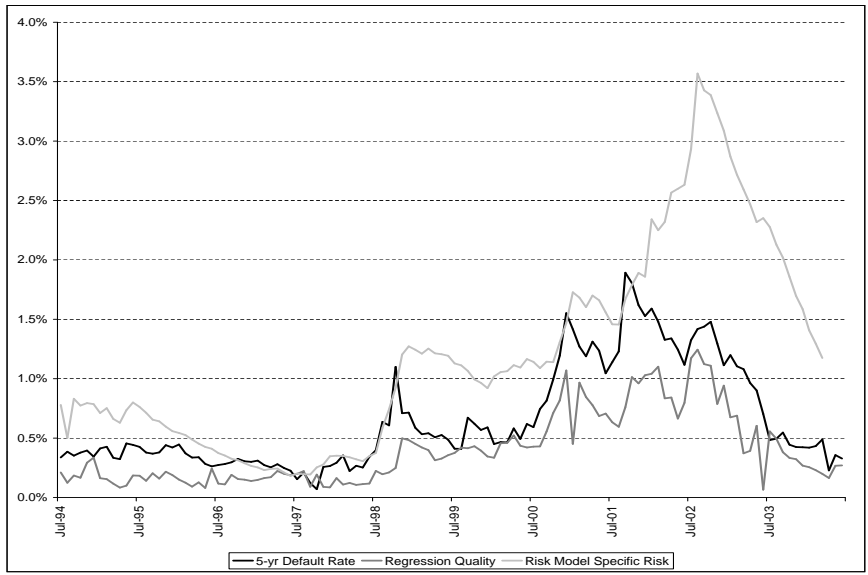}%
\caption{{\small Implied Default Probability, Average Pricing Error of the Regression, and Risk Model Specific Risk, A-rated Industrials.}}%
\label{fig:estimation-stats}%
\end{figure}

Figure \ref{fig:estimation-stats} demonstrates the results of the estimation procedure for the A-rated Industrials, performed monthly for 10 years from July 1994 until June 2004, using the end-of-month prices of senior unsecured bonds in the Lehman Brothers credit database. We show the time series of the 5-year annualized default probability versus the weighted average pricing error of the cross-sectional regression. The latter is defined as the square root of the objective function given by the equation (\ref{eq:survival-regression-OF}), with weights normalized to sum up to 1. 

The regression quality has tracked the level of the implied default rates - the higher implied default rates are associated with greater levels of idiosyncratic errors in the cross-sectional regression. This pattern is consistent with the assessment of the issuer-specific excess return volatility during the same period given by the Lehman Brothers multi-factor risk model (see \cite{LehmanRiskModel}). We show for comparison the exponentially-weighted specific risk estimates for the A-rated Basic Industries bucket.  

As a final remark we would like to note that the choice of recovery rates used in our model is obviously very important. After all, the main impetus for this methodology was the recognition that recovery rates are a crucial determinant of the market behavior for distressed bonds. Both the cross-sectional, i.e. industry and issuer dependence, and the time-series, i.e. business cycle dependence of the recovery rates is very significant (see \cite{LossCalc}, \cite{Altman-2004}, \cite{Altman-2005}). Nevertheless, it is often sufficient to use an average recovery rate, such as 40\% which is close to the long-term historical average across all issuers, for the methodology to remain robust across the entire range of credit qualities.  

In a more ambitious approach, the recovery rate can be estimated by a second-stage likelihood maximization after obtaining the best fit of the exponential spline coefficients given a recovery value as a parameter. This would yield best fit or `market implied' recovery rates. Presumably, the industry dependence can also be handled by introducing different 'implied' recoveries for different industries, assuming the number of independently priced bonds is sufficiently large to maintain statistical significance of the obtained results. We have found, however, that the majority of investment grade bonds do not efficiently price the recovery and therefore this program, while theoretically possible, is difficult to implement in practice.

\section{Issuer and Sector Credit Term Structures} \label{sec:issue_sector_ts}

Having estimated the term structure of survival probabilities, we can now define a set of valuation, risk and relative value measures applicable to collections of bonds such as those belonging to a particular issuer or sector, as well as to individual securities. In practice, to preserve the consistency with market observed bond prices encoded in the survival probability, we define the issuer and sector credit term structures for the same set of bonds which were used in the exponential spline estimation procedures. 

As discussed earlier, the conventional Z-spreads are not consistent with survival-based valuation of credit risky bonds. The same can be said about the yield spreads, I-spreads, asset swap spreads, durations, convexities and most other measures which investors currently use day to day. Ultimately, this inconsistency is the source of the breakdown of the conventional spread measures in distressed situations. The market participants know this very well and they stop using these measures for quoting or trading distressed bonds. This situation is commonly referred as {\em bonds trading on price}.

In this Section we will define a host of measures which are consistent with the survival-based approach. We note, however, that the definitions presented in this section do not depend on the specific choice of the exponential splines methodology for fitting survival probability term structures. They can be used in conjunction with any term structure of survival probabilities which is consistent with reduced-form pricing methodology assuming fractional recovery of par - for example one calibrated to the CDS market.

\subsection{Survival and Default Probability Term Structures} \label{sec:survival_prob_ts}

The survival probability term structure is a direct output from the empirical estimation process described in the previous section. Once we have estimated the spline coefficients $\beta_{k}$ and the long-term decay parameter $\eta$, the survival probability is defined by the equation (\ref{eq:survival-spline}).

\begin{figure}[t]
\includegraphics[height=3in,width=4.5in]{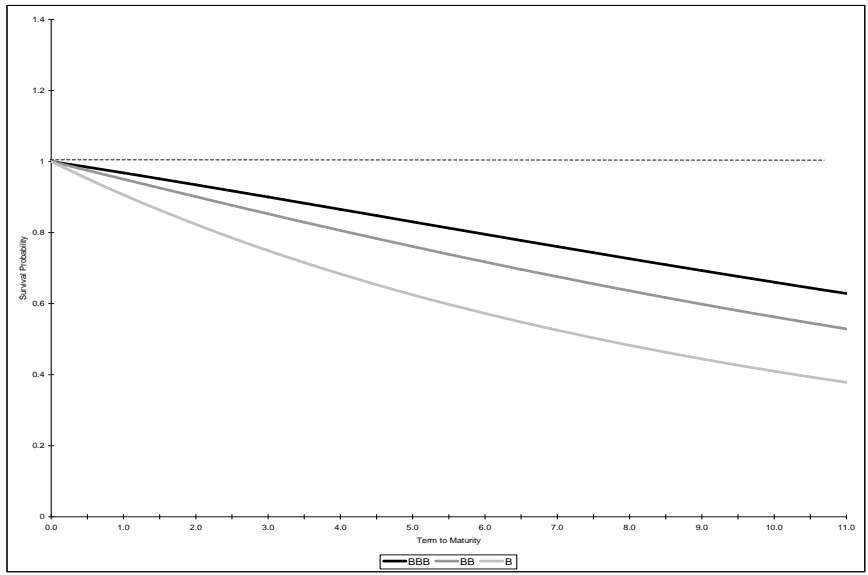}%
\caption{{\small Survival probability term structures for different credit risk levels.}}%
\label{fig:surv-prob}%
\end{figure}

Correspondingly, the cumulative default probability is defined as:

\begin{equation} \label{eq:ts-cum-default-prob}
D(t) = 1 - Q(t)
\end{equation}
which one can recognize as a special case of eq.\ (\ref{eq:default-prob-finite}). Figure \ref{fig:surv-prob} shows the shapes of the survival probabilities for credit sectors with varying risk levels, from BBB-rated to B-rated credit.

\subsection{Hazard Rate and ZZ-Spread Term Structures} \label{sec:hazard_zz_ts}

Credit investors and market practitioners have long used definitions of spread which correspond to the spread-discount-function methodology, outlined in Section \ref{sec:conventional}. The most commonly used measures, Z-spread and OAS, explicitly follow the discounting function approach. Others, such as yield spread or I-spread, implicitly depend on bond-equivalent yields which in turn follow from discounting function approach. Thus, all of these measures neglect the dependence of the bond price on the recovery value and the debt acceleration in case of default. Therefore, these measures become inadequate for distressed bonds. 

In the survival-based approach, spreads are not a primary observed quantity. Only the prices of credit bonds have an unambiguous meaning. Spreads, however we define them, must be derived from the term structure of survival probabilities, fitted to the bond prices. 

There is only one spread measure which is defined directly in terms of survival probabilities. It corresponds to the spread of a hypothetical credit instrument which pays \$1 at a given maturity, pays no interest and pays nothing in case of default. The continuously compounded zero-recovery zero-coupon ZZ-yield for such a bond is defined as:

\begin{eqnarray} \label{eq:ts-ZZ-yield}
e^{-Y_{ZZ} \, T} & = & Q(t) \, Z_{base}(T) \nonumber \\
Y_{ZZ}(T) & = & -\frac{1}{T} \log \left( Q(t) \, Z_{base}(T) \right) 
\end{eqnarray}

Recall that the corresponding continuously compounded risk-free zero-coupon rate is defined as (\ref{eq:zc_yield_curve}).

\begin{equation} \label{eq:ts-rf-yield}
y(T) = -\frac{1}{T} \log \left( Z_{base}(T) \right) 
\end{equation}

From these two definitions, we obtain the continuously compounded ZZ-spread as the difference between the ZZ-yield and risk-free zero-coupon rate:

\begin{equation} \label{eq:ts-ZZ-spread}
S_{ZZ}(T) = Y_{ZZ}(T) - y(T) = -\frac{1}{T} \log \left( Q(t) \right) 
\end{equation}

Substituting the definition of the survival probability, we see that the ZZ-spread is equal to the average hazard rate for the maturity horizon of the hypothetical zero-recovery zero-coupon credit bond.

\begin{equation} \label{eq:ts-ZZ-hazard}
S_{ZZ}(T) = \frac{1}{T} \int_{0}^{T} h(s) \, ds 
\end{equation}

Equivalently, we can say that the instantaneous forward ZZ-spread is identically equal to the hazard rate of the issuer. The hazard rate also has a meaning of instantaneous forward default probability, i.e. the probability intensity of default during a small time interval in the future provided that the issuer has survived until that time. Thus, the forward ZZ-spread is equal to the forward default intensity:

\begin{equation} \label{eq:ts-ZZ-spread-hazard}
S^{fwd}_{ZZ}(t) = h(t)
\end{equation}

This result coincides with the conclusions in \cite{Duffie-1998} regarding the hazard rate in reduced form models under the FRP recovery assumption. The novelty of our approach is that we do not estimate the hazard rate from spread curves, but conversely, derive the spread curves from hazard rates that are obtained directly from the bond prices via the fitted survival probability. Since the conventional spread estimates are not consistent with the survival-based approach, deriving the ZZ-spread or the hazard rates from such spread measures is fraught with inaccuracies and biases, especially for high hazard rates. Only if the recovery rate is equal to zero and for the case of zero coupon credit bonds the conventional Z-spread becomes equal to the forward ZZ-spread and becomes consistent with the survival-based valuation - in concordance with our earlier assertions.

Using the exponential spline representation of the survival probability term structure (\ref{eq:survival-spline}), we can derive the hazard rate term structure as follows:

\begin{equation} \label{eq:ts-hazard-spline}
h(t) = -\frac{d}{dt} \log Q(t) = \frac{\sum_{k=1}^{3}{k \, \eta \, \beta_{k} \, e^{-k \, \eta \, t}}}{\sum_{k=1}^{3}{\beta_{k} \, e^{-k \, \eta \, t}}}
\end{equation}
where the constraint (\ref{eq:survival-regression-constraint2}) guarantees the positivity of the hazard rates. 

\begin{figure}[t]
\includegraphics[height=3in,width=4.5in]{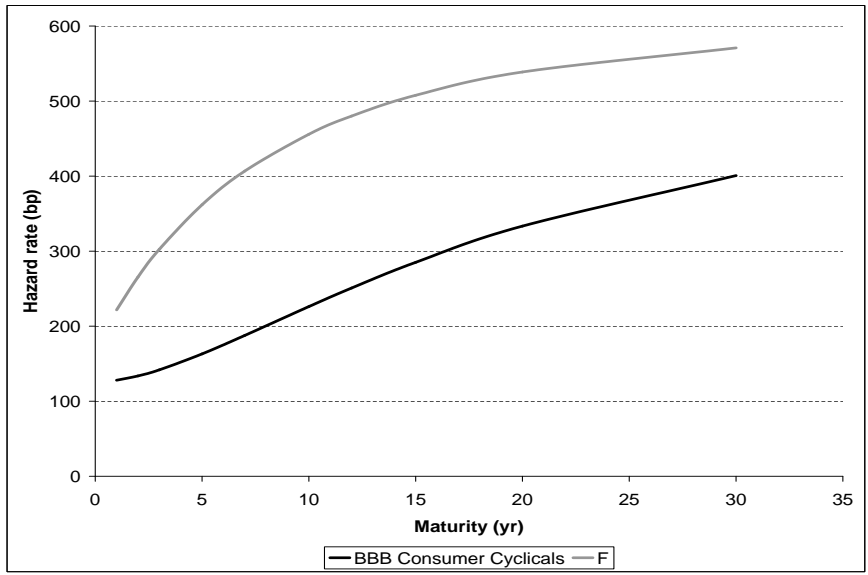}%
\caption{{\small Hazard rates (forward ZZ-spreads) for Ford and BBB Consumer Cyclicals, as of 12/31/03.}}%
\label{fig:zz-spreads}%
\end{figure}

Figure \ref{fig:zz-spreads} shows the result of estimation of the hazard rate (forward ZZ-spread) term structures for Ford and for the BBB Consumer Cyclicals sector. 

Even though we classify the hazard rate primarily as a valuation measure, it can be used for relative value assessments. It is important to remember that the implied hazard rate does not correspond to an actual forecast. Like other market-implied parameters, it also incorporates in a complex way a host of risk premia which reflect both credit and non-credit factors such as recovery rate risk, liquidity, etc. Nevertheless, fitted hazard rate term structures may provide valuable clues beyond the conventional spread analysis. 

For example, the Ford curve in Figure 4 is substantially wider than the sector curve, signalling the higher credit risk associated with Ford bonds. The Ford curve has a distinctive shape, with the maximum differential to sector curve in intermediate maturities, suggesting that these maturities offer the best opportunities for monetizing views on the relative risk and return between this issuer and the industry sector.

\subsection{Par Coupon and P-Spread Term Structures} \label{sec:p_spread_ts}

While the definition of the ZZ-spread presented in the previous subsection is quantitatively sound, it is not likely to be of much value for practitioners because the zero-coupon zero-recovery credit bonds do not actually exist in the marketplace. There exist pure discount (zero-coupon) securities, particularly in the short term CP market, but all of them are subject to equal-priority recovery rules and therefore cannot be considered zero-recovery. On the other hand, there exist credit derivatives such as digital default swaps which can have a contractual zero recovery, but they do have premium payments and therefore cannot be considered as an equivalent of a zero-coupon bond. Therefore, the usefulness of the ZZ-spread as a relative value measure is limited - one can obtain from it some insight about the issuer or the sector but not about a particular security.

The vast majority of the credit market consists of interest-bearing instruments subject to equal-priority recovery in case of default. A practically useful relative value measure should refer to such instruments and should contrast them with credit risk-free instruments such as Treasury bonds or interest rate swaps which provide a funding rate benchmark . One such measure is the fitted par coupon - i.e. a coupon of a hypothetical bond of a given maturity which would trade at a par price if evaluated using the fitted issuer survival probability term structure. Correspondingly, the par spread is defined by subtracting the fitted par yield of risk-free bonds from the fitted par coupon of credit-risky bonds. 

One must note that the par coupon and par spread measures do not correspond to a specific bond - these are derived measures based on the issuer survival curve and a specific price target equal to par, i.e. 100\% of face value. The par price of the bond has a special significance, because for this price the expected price return of a risk-free bond to maturity is precisely zero. Therefore, the par yield of a risk-free bond reflects its expected (annualized) total return to maturity and the par spread of a credit-risky bond reflects its (risky) excess return to maturity. Thus, the par spread defined above can be considered a consistent (fair) relative value measure for a given issuer/sector for a given maturity horizon. 

For example, if two different issuers wanted to buy back their outstanding bonds in the secondary market and instead issue new par bonds of the same maturity $T$ then the fair level of the coupons which the market should settle at, assuming no material change in the issuers' credit quality, would be given by the respective fitted par coupons. 
Correspondingly, investors considering these new bonds can expect excess returns to maturity T equal to their respective fitted par spreads. If one of those spreads is greater than the other - this would represent a relative value which the investors should contrast with their views of the issuers' credit risks to the said maturity horizon.
For bonds with coupon frequency $q$ (usually annual $q=1$, or semi-annual $q=2$) with an integer number of payment periods until maturity $t_{N}=N/q$ we define the par coupon term structure by solving for the coupon level from the pricing equation [11]. 

\begin{equation} \label{eq:par-coupon-survival}
	C_{P}(t_{N}\left|q\right.) = q \, \frac{1 - Z(t_{N}) \, Q(t_{N}) - R \, \sum_{i=1}^{N}{ Z(t_{i}) \, \left( Q(t_{i-1})  - Q(t_{i}) \right) }}{\sum_{i=1}^{N}{ Z(t_{i}) \, Q(t_{i}) + \frac{1}{2} R \, \sum_{i=1}^{N}{ Z(t_{i}) \, \left( Q(t_{i-1})  - Q(t_{i}) \right) }}}
\end{equation}

Contrast this definition with what the one that would be consistent with spread discount function based approaches:

\begin{equation} \label{eq:par-coupon-discount}
	C_{disc}(t_{N}\left|q\right.) = q \, \frac{1 - Z_{base}(t_{N}) \, Z_{spread}(t_{N})}{\sum_{i=1}^{N}{ Z_{base}(t_{i}) \, Z_{spread}(t_{i}) }}
\end{equation}

We can see that the latter definition only coincides with the former if we assume that the recovery rate is zero and that the spread discount function is equal to the survival probability.

\begin{figure}[t]
\includegraphics[height=3in,width=4.5in]{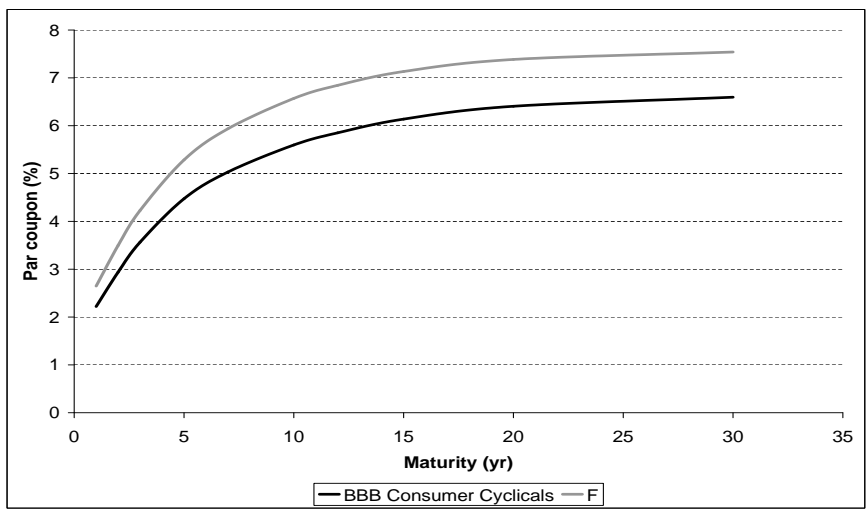}%
\caption{{\small Fitted par coupon for Ford and BBB Consumer Cyclicals, as of 12/31/03.}}%
\label{fig:fitted-par-coupon}%
\end{figure}

An example of the fitted par coupon term structure is shown in Figure \ref{fig:fitted-par-coupon} where we show our estimates for Ford and BBB Consumer Cyclicals sector as of 12/31/2003. Note that the shape of these fitted curves - very steep front end and flattish long maturities - is largely determined by the shape of the underlying risk-free curve (we used the swaps curve in this case), with the credit risk being a second-order modification for most issuers and sectors, except those that trade at very deep discounts.
 
Let us also define the par coupon of the risk-free bond in a similar fashion:

\begin{equation} \label{eq:par-yield-riskfree}
	C_{P}^{base}(t_{N}\left|q\right.) = q \, \frac{1 - Z_{base}(t_{N})}{\sum_{i=1}^{N}{ Z_{base}(t_{i})}}
\end{equation}

The par spread (P-spread) to the base curve (either Treasury or swaps) can then be derived by subtracting the par base yields from the par risky coupons of the same maturities:

\begin{equation} \label{eq:par-spread}
	S_{P}(T\left|q\right.) = C_{P}(T\left|q\right.) - C_{P}^{base}(T\left|q\right.)
\end{equation}

\begin{figure}[t]
\includegraphics[height=3in,width=4.5in]{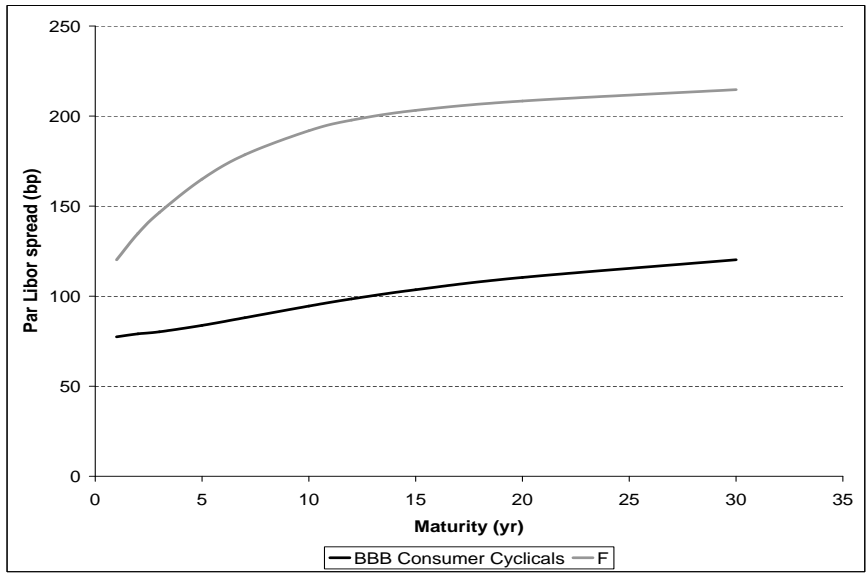}%
\caption{{\small Fitted Libor P-spread for Ford and BBB Consumer Cyclicals, as of 12/31/03.}}%
\label{fig:fitted-p-spread}%
\end{figure}

Figure \ref{fig:fitted-p-spread} demonstrates the fitted par Libor spread term structures using the same example of the Ford and BBB Consumer Cyclicals sectors.

\subsection{Constant Coupon Price (CCP) Term Structure} \label{sec:ccp_ts}

Since we argued that the price-based estimation techniques are more consistent than those fitting yields or spreads, it is useful to define a set of credit term structures expressed in terms of bond prices. For any integer number of payment periods  , we define the constant coupon price (CCP) term structure as the price level of the bond with a pre-set coupon (e.g., coupons of 6\%, 8\%, 10\%).

\begin{eqnarray} \label{eq:CCP}
	P(t_{N}\left|C\right.) & = & Z_{base}(t_{N}) \, Q(t_{N}) + \frac{C}{q} \sum_{i=1}^{N}{ Z_{base}(t_{i}) \, Q(t_{i}) } \nonumber \\
	      & + & R \, \left(1 +  \frac{C}{2q} \right) \sum_{i}^{N}{ Z_{base}(t_{i}) \, \left(Q(t_{i-1}) - Q(t_{i})\right) }
\end{eqnarray}

\begin{figure}[t]
\includegraphics[height=3in,width=4.5in]{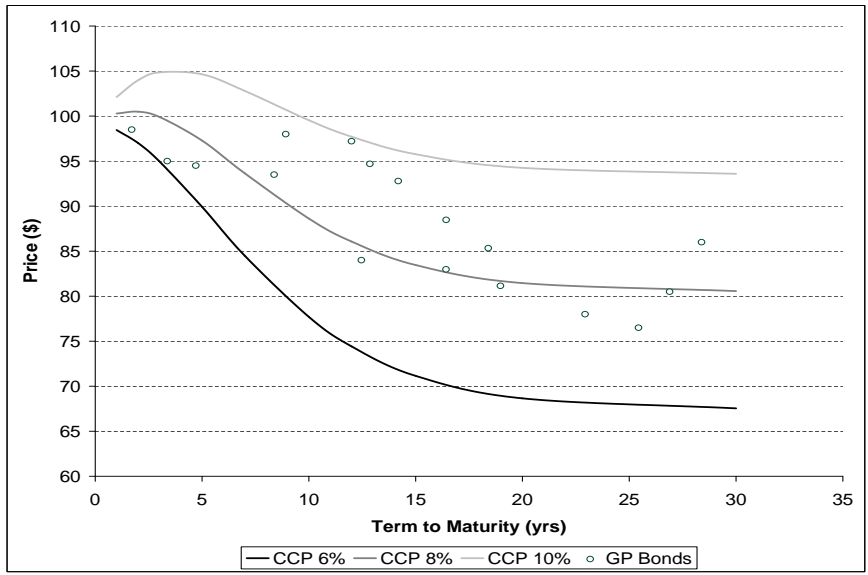}%
\caption{{\small CCP term structures and bond prices, Georgia Pacific, as of 12/31/03.}}%
\label{fig:fitted-ccp}%
\end{figure}

Figure \ref{fig:fitted-ccp} shows estimated CCP term structures for Georgia Pacific as of December 31, 2002. The prices are calculated as fractions of a 100 face, i.e. par price 100\% appears as 100. We observe that the 6\%, 8\% and 10\% Constant Coupon Price term structures neatly envelope the scatterplot of prices of individual bonds which have coupon levels ranging from 6.625\% to 9.625\%. Importantly, all three CCP term structures correspond to the same term structure of survival probability, and thus embody the same credit relative value. To the extent that a price of a bond with a particular fixed coupon is in line with the level suggested by the CCP term structures, this bond also reflects the same credit relative value. Thus, a graph like this can serve as a first crude indication of the relative value across bonds of a given issuer or sector - especially when there are securities whose prices are substantially different from the corresponding fitted price curve levels. In the next section we develop a more precise measure for assessing such relative value, which we call Default Adjusted Spread. 

Note also how the CCP term structures tend to become flat at longer maturities - this is a reflection of the fact that Georgia Pacific was trading at elevated levels of the implied default risk, with the fitted hazard rate exceeding 1000 bp at maturities longer than 5 years. At such levels of credit risk the implied survival probability to 10 years or longer is only about 35\%, and the recovery scenario at longer maturities becomes the dominant one. This, in turn, leads to a flat term structure of prices. 

One might say that the reason for this is that the high probability of an early default scenario causes the bonds to trade with an effective life much shorter than their nominal maturity. We show in the next section that the properly defined duration measure for credit bonds that is consistent with the survival-based valuation will exhibit the same feature, with the duration becoming much shorter for bonds with higher levels of the implied credit risk.

\subsection{Bond-Implied CDS (BCDS) Term Structure} \label{sec:BCDS}

The survival-based valuation approach is well suited for the CDS market. In fact it has been the market practice since its inception. By deriving the bond-implied CDS spreads within the same framework we are aiming to give investors an apples-to-apples relative value measure across the bond and CDS markets. The pricing relationship for credit default swaps was defined in (\ref{eq:pricing-CDS}). The only difference here is that rather than observing the CDS market and calibrating the survival probability to it, we derive the bond-implied CDS (BCDS) from the previously fitted values of the survival probability term structure. 

\begin{equation} \label{eq:BCDS}
	S_{BCDS}(t_{N}) = 2 q \, (1-R) \, \frac{\sum_{i=1}^{N}{ Z_{base}(t_{i}) \, \left( Q(t_{i-1}) - Q(t_{i}) \right) }}{\sum_{i=1}^{N}{ Z_{base}(t_{i}) \, \left( Q(t_{i-1}) + Q(t_{i}) \right) }}
\end{equation}

Another difference of BCDS from bond-like spread measures such as P-spread, is that for BCDS we use the quarterly coupon frequency $q=4$, rather than more common semi-annual $q=2$ for bonds. The BCDS term structure gives yet another par-equivalent measure of spread for credit-risky issuers in addition to the P-spread. The base discounting function $Z_{base}(t_{i})$ must be taken typically to be LIBOR or another funding-related function in order for the BCDS spread to have a meaning of excess return over cash. In section \ref{sec:cds_bond_basis} we will show that there exists a complementarity between the BCDS term structure and the properly defined credit-risk-free benchmark security, proving that the BCDS spread is a clean measure of excess return, free of biases associated with non-par prices. 

\subsection{Forward Spreads and Forward Trades} \label{sec:fwd_spread_ts}

The BCDS spread definition is also the best starting point to define the forward credit spread measure consistent with survival-based methodology. Since most other measures rely on the comparison between the credit-risky and risk-free bonds, they would suffer from inherent biases if naively extrapolated to forward space. The forward prices are well defined only for zero coupon riskless bonds, in which case the forward price is simply the price of final maturity bond expressed in units of bond maturing at the forward pricing date: $Z(t,T) = Z(T)/Z(t)$. However, a similar comparison is meaningless for coupon-bearing bonds, and especially if the bonds are credit-risky. Therefore, the spread and yield measures defined for coupon-bearing bonds do not lend themselves well as a starting point for forward spread calculation.

In contrast, the BCDS spread is not defined with respect to any underlying bond -- it is instead defined purely as a function of the survival probability and base discount function. Both of these quantities have a well defined meaning in forward space:

\begin{eqnarray} 
Z^{fwd}_{base}(t,T) & = & E_{0}\left\{ e^{-\int_{t}^{T}{r_{s} ds}} \right\} = e^{-\int_{t}^{T}{f_{s} ds}} \label{eq:fwd-base-discount} \\
Q^{fwd}(t,T)  & = & E_{0}\left\{ 1_{T < \tau} \left| t < \tau \right. \right\} = e^{-\int_{t}^{T}{h_{s} ds}} \label{eq:fwd-survival-prob}
\end{eqnarray}
where the interpretation of the forward survival probability is as the conditional survival probability until time $T$, under the condition that the default has not occured until time $t$.

Therefore, by simply replacing the current observed discount and survival functions in (\ref{eq:BCDS}) with their forward counterparts estimated for the forward starting time $t$, one obtains the forward BCDS spread for time $t$ and forward tenor $T=t_{N}$.

Let us now contrast this calculation with more market-driven notions of a forward CDS contract and a forward CDS trade (whether or not we are talking about the actual CDS or a hypotethical bond-implied CDS is not important at this point).

A forward CDS contract corresponds to buying or selling credit protection that is active for a period of time in the future at a premium whose level is preset today, but payable only during that future period. For example, a 5 year CDS 2 years forward, referred to as 2x5 forward CDS, provides protection which starts in year $T_1=2$ and continues for $T=5$ years afterward, ending in year $T_2=T_1+T=7$ from today, with premiums paid quarterly during that period. Importantly, the forward CDS knocks out and provides no protection if a credit event occurs prior to the forward start date. 

While the market participants rarely trade such a contract, they often trade long-short pairs of spot CDS which are almost (but not exactly) identical to a forward CDS. These pair trades are known in the market as CDS steepeners and flatteners. For example, buying a 2x5 forward CDS protection is similar to a 2s-7s curve steepener trade whereby one sells protection to 2 year maturity and buys equal notional protection to 7 year maturity. This trade would outperform if the spread curve steepens, hence the name {\em steepener}. The opposite trade, which would outperform if the spread curve flattens, is correspondingly known as a {\em flattener}.

In terms of protection payment leg such a trade is indeed equivalent to a forward CDS - if the credit event occurs before $T_1$ the long and the short protection payments will cancel each other, while if the credit event occurs between the year $T_1$ and $T_2$ only the long protection leg is active and it will provide the same protection payment as the forward CDS contract.

The premium legs are somewhat different between the forward CDS and the long-short pair trade. Table \ref{tab:ForwardCDS} demonstrates the premium cash flows for each of the trades. As we can see, the steepener pair trade potentially has non-zero premium payments during the first two years (unless the CDS term structure is flat), while the forward CDS does not. Correspondingly, the net premium paid on a steepener trade between $T_1$ and $T_2$ will also be different from the forward CDS premium in this example. 

\begin{table}[tp]
\begin{tabular}{|l|c|c|c|c|c|c|} \hline
Cash Flow Time 							& $t=0$		& \multicolumn{2}{|c|}{$[0 < t \le T_1]$} 	& \multicolumn{2}{|c|}{$[T_1 < t \le T_2]$} & $t=T_2$ 		\\ \hline
Cash Flow Type 							& Upfront	& Premium				& Default		& Premium								& Default 				& Maturity 	\\ \hline \hline
Long FwdCDS [$T_1$ x $T$]		& $0$			& $0$						&	$0$				&	$-S_{fwd}$						&	$1-R$						&	$0$								\\ \hline \hline
Short CDS [$T_1$]				 		& $0$			& $S_1$					&	$-(1-R)$	&	$0$										&	$0$							&	$0$							\\ \hline
Long CDS [$T_2=T_1+T$]			& $0$			& $-S_2$				&	$1-R$			&	$-S_2$		 						&	$1-R$						&	$0$								\\ \hline \hline
Hedged Fwd									& $0$			& $S_2-S_1$			&	$0$				&	$S_2-S_{fwd}$					&	$0$							&	$0$								\\ \hline
\end{tabular}
	\caption{Cash Flows of Forward CDS hedged by a CDS flattener}
	\label{tab:ForwardCDS}
\end{table}

Since the forward CDS hedged by a long-short CDS pair provides no protection payment under any default scenario, the present value of the protection leg of such a trade is identically zero. Therefore, we must require that the net premium cash flows of a hedged forward CDS trade also have zero present value. The present value of each basis point paid from today until the earlier of a given maturity or the occurrence of the credit event is given by the risky PV01 $\pi(T)$ to corresponding maturity (\ref{eq:rpv01-CDS}). Hence, this requirement can be written as: 

\begin{equation} \label{eq:FwdCDS-pricing}
	(S_2-S_1) \, \pi_1 + (S_2-S_{fwd}) \, \left(\pi_2 - \pi_1\right) = 0
\end{equation}

Here,  $S_1$ is the spot CDS spread for the starting date, $S_2$  is the spot CDS spread for the ending date, and $S_{fwd}$ is the forward CDS spread for the given interval of maturities. $\pi_1$  and $\pi_2$  are the risky PV01s for the starting and ending dates, respectively. The first term represents the present value of the stream of cash flows between today and the starting date, and the second term corresponds to the present value of cash flows between the starting and ending dates.

From this condition we obtain the following simple relationship for forward CDS spreads:

\begin{eqnarray} \label{eq:FwdCDS-spread}
	S_{fwd} & = & \frac{S_2 - \kappa \, S_1}{1 - \kappa} = S_2 + \frac{\kappa}{1-\kappa} (S_2 - S_1) \\
	\kappa & = & \frac{\pi_1}{\pi_2} \nonumber 
\end{eqnarray}

Thus, the forward CDS spread is equal to a weighted average of the spot CDS spreads to initial and final maturity, with the weights determined by the ratio of risky PV01s to each maturity (note that $0<\kappa<1$). If the CDS curve is upward sloping ($S_2>S_1$), then the forward CDS spread is higher than the long maturity spot spread. If the CDS curve is inverted ($S_2>S_1$), then the forward CDS spread is lower than the long maturity spot spread.

As a relative value measure, the forward spread of the issuer can be a powerful tool for investors seeking to choose the best exposure term for the given credit. Market segmentation plays an important role in driving the shapes of credit term structures, and it can lead at times to unusually rich or cheap segments of the term structure. Comparing the forward spreads across term structure segments is the most consistent way to zoom in on such relative value opportunities.

\section{Bond-Specific Valuation Measures} \label{sec:bond_specific}

So far we have developed a set of term structures which encode our knowledge about the issuer (or sector) as a whole, rather than about the specific bond. In particular, our primary measure, the term structure of survival probability, clearly refers to the issuer and not to any particular bond issued by this issuer. It would make no sense to say that the XYZ 6.5\% bond maturing in 10 years has a term structure of survival probabilities, but it does make sense to say that the XYZ issuer has a term structure of survival probabilities which was fitted using the price of the above mentioned bond, as well as other bonds of the same issuer, if available.

When it comes to a particular bond, investors are typically concerned with their fair value and relative value with respect to other bonds of the same issuer or sector. The estimate of the fair value for a given bond is a straightforward application of the issuer- or sector-specific fair value metrics to the particular maturity and coupon of the security under investigation. The answer to the second question lies in the comparison of the market-observed bond price with the estimated fair value price. The Default-Adjusted Spread (DAS) measure, introduced below, provides an unambiguous and consistent way to make such a comparison, free of biases associated with the term to maturity or level of coupon, which plague the conventional spread measures. 

\subsection{Fitted Price and P-Spread} \label{sec:fitted_price_spread}

The CCP measure introduced in the previous section determines the precise fit of the bond's price for a hypothetical bond with a pre-set coupon and maturity chosen so that there are integer number of payment periods and no accrued coupon amount as of pricing date. 

We can easily extend this generic fitted price measure to any given bond by defining it as the clean price such a bond would have if it was priced precisely by the issuer- (or sector) fitted survival probability term structure. Clearly, the notion of the fitted price depends on the context of the fit, i.e. whether we are talking about the fair value with respect to issuer or sector. A bond can be undervalued with respect to other bonds of the same issuer, but overvalued with respect to majority of the bonds in the larger industry or rating sector.

Since the accrued interest $A_{int} = C \, t_{acc}$ is a known value depending on the coupon level and time accrued since issue or last coupon $t_{acc}$, the fitted price in our implementation is precisely equal to the market price less the regression residual from the cross-sectional estimation (\ref{eq:survival-regression}). 

Outside of the regression context, one would calculate it using the term structure of the `fair' survival probability for the issuer, such as the one calibrated from the benchmark CDS spread levels: 

\begin{eqnarray} \label{eq:bond-fitted-price}
	P^{fit} & = & Z_{base}(t_{N}) \, Q(t_{N}) + \frac{C}{q} \sum_{i=1}^{N}{ Z_{base}(t_{i}) \, Q(t_{i}) } \nonumber \\
	      & + & R \, \left(1 + \frac{C}{2q} \right) \, \sum_{i}^{N}{ Z_{base}(t_{i}) \, \left(Q(t_{i-1}) - Q(t_{i})\right) } \nonumber \\
	      & - & C \, t_{acc}
\end{eqnarray}

The fitted par coupon of a given bond is defined as the coupon which would make this bond's clean price equal to par when evaluated using the suitably chosen fitted survival term structure. For a given maturity date, we must modify equation (\ref{eq:par-coupon-survival}) for the par coupon to account for the effect of the non-current coupon and the non-zero accrued interest amount: 

\begin{equation} \label{eq:bond-par-coupon}
	C_{P}^{fit} = q \, \frac{1 - Z_{base}(t_{N}) \, Q(t_{N}) - R \, \sum_{i=1}^{N}{ Z_{base}(t_{i}) \, \left( Q(t_{i-1})  - Q(t_{i}) \right) }}{
	\sum_{i=1}^{N}{ Z_{base}(t_{i}) \, Q(t_{i}) + \frac{1}{2} \,R \, \sum_{i=1}^{N}{ Z_{base}(t_{i}) \, \left( Q(t_{i-1})  - Q(t_{i}) \right) } } - t_{acc}}
\end{equation}	 

The difference between the fitted par coupon and the correspondingly defined fitted par base (LIBOR or Treasury) rate for the same maturity could be termed the fitted (or fair) P-spread. 

\begin{equation} \label{eq:bond-P-spread}
	S_{P}^{fit} = C_{P}^{fit} - C_{P}^{base}
\end{equation}	  

This spread, however, does not correspond to the observed price of the bond. It is instead a P-spread that the bond would have if it was priced precisely, without any residual errors, by the corresponding issuer- or sector-specific fitted survival probability term structure. Next, we turn our attention to pricing errors and relative value measures.

\subsection{Default-Adjusted Spread and Excess Spread} \label{sec:DAS}

For a long time credit investors have been using spread measures such as nominal spread, I-spread, OAS or Z-spread, to assess the relative value across various bonds of the same issuer or sector (see O'Kane and Sen [2004] for a glossary of terms). In the previous sections we have demonstrated that these measures have inherent biases because all of them rely on the strippable cash flow valuation assumption which is inadequate for credit-risky bonds. 

We argued that spread-like measures which can be interpreted in terms of risky excess return to maturity and are in agreement with the survival-based valuation, correspond to idealized par bonds or par-equivalent instruments such as CDS. However, bonds in the secondary market typically trade away from par. It is important to measure the degree by which the bond's price deviates from the ``fair'' price corresponding to its coupon level and the term structure of the underlying interest rates. As explained in the previous subsection, the latter is the analogue of the constant coupon price term structure. For example, in terms of credit relative value a 6\% bond trading at a price close to the 6\% CCP curve is not any different from a 10\% bond trading close to 10\% CCP curve. On the other hand, if the first bond were trading above the 6\% CCP price while the second was trading below the 10\% CCP price, we would say that the first bond is rich and the second bond is cheap, even though the observed price of the first bond might be less than that of the second bond.   

Fortunately, our estimation methodology for survival probability term structures lends itself naturally to a robust determination of the rich/cheap measures as described above. Indeed, we impose a fairly rigid structural constraint of the shape of the survival probability term structure by adopting the exponential splines approximation. The result was that the cross-sectional regression which gives the spline coefficients does not in general price any of the bonds precisely. Instead, we make a trade-off between the individual bond pricing precision and the robustness of the overall fit. The bond-specific pricing errors from the best fit of the survival probability term structure (i.e. cross-sectional regression residuals $\epsilon_{reg}$) then become a natural candidate for the relative value across the bonds - if the residual is positive we say that the bond is rich, and if the residual is negative we say that it is cheap.

We can express the pricing error in terms of a constant Default-Adjusted Spread (DAS) which acts as an additional discount (or premium) spread that replicates the bond's present value:

\begin{eqnarray} \label{eq:bond-das}
	P^{mkt} & = & P^{fit} + \epsilon_{reg}  \nonumber \\
	 & = & Z_{base}(t_{N}) \, Q(t_{N}) \, e^{-DAS \, t_N} + \frac{C}{q} \sum_{i=1}^{N}{ Z_{base}(t_{i}) \, Q(t_{i})  \, e^{-DAS \, t_i}} \nonumber \\
	      & + & R \, \left(1 + \frac{C}{2q} \right) \, \sum_{i}^{N}{ Z_{base}(t_{i}) \, \left(Q(t_{i-1}) - Q(t_{i})\right)  \, e^{-DAS \, t_i} } \nonumber \\
	      & - & C \, t_{acc} 
\end{eqnarray}

Here, $P_{mkt}$ is the observed clean market price, and $A_{int} = C \, t_{acc}$ is the known accrued interest. Comparing this equation to (\ref{eq:bond-fitted-price}) one can see that if $P_{mkt}=P_{fit}$ then $DAS=0$. Default-Adjusted Spread should be interpreted as a bond-specific premium/discount which reflects both market inefficiencies such as liquidity premia and biases related to persistent market mispricing of credit bonds. We say that a positive DAS (negative regression residual $\epsilon_{reg}$) signals cheapness and the negative DAS (positive residual $\epsilon_{reg}$) signals richness across the bonds of the same issuer. This is in line with the usual meaning assigned to spreads. Moreover, we can say that the DAS differential for two bonds reflects the `clean' relative value between them.

Having determined this last piece of the puzzle, we can now define the excess spread $S_{X}$ of the credit bond as the previously derived fitted P-spread $S_{P}$ plus the default-adjusted spread. Since both the fitted P-spread and DAS have a clear relative value interpretation, this measure of the total spread continues to have a  meaning as a measure of excess return over the credit risk-free bonds or swaps curve, whichever is used as a base curve (hence the name {\em excess spread}).

\begin{equation} \label{eq:bond-excess-spread}
	S_{X} = S_{P} + DAS
\end{equation}	  

It is clear from its definition that the excess spread measure should not be directly used in any discounting calculation, as it contains the P-spread measure that is not related to observed cash flows of the bond. This highlights one more time the fact that consistent relative value measures for credit bonds are not compatible with naive (strippable) discounted cash flow methodology.



Table \ref{tab:DAS} shows the calculated values for Calpine bonds as of 6/30/2003. Note the large differences between the Z-spread and P-spread measures of the bonds across all maturities, ranging from 250 to 400 bp. Note also that looking at Z-spreads alone it would be difficult to the relative value between different bonds. By contrast, DAS provides a clean measure of such relative value. It appears that most high-coupon bonds are rich (have negative DAS), while the low-coupon bonds are cheap. Such patterns of relative value driven by the coupon levels are quite often observed, and are a consequence of the market mispricing driven by the use of conventional spread measures. 

\begin{table}[tp]
\begin{tabular}{|l|c|c|c|c|c|c|c|c|} \hline
Description					& Maturity 				& Coupon				& Z-spread		& P-spread		& DAS				& Price 	& Fitted 			 	& Price 				 	\\ 
										& (yrs)						& 							& 						& 						& 					& 			 	& Price 				& Residual 				\\ \hline \hline
CPN $8\frac{1}{4}$ 8/05			& 2.13						& 8.25					&	1649				&	1949				&	-68				&	82.00		& 81.02					& 0.98						\\ \hline 
CPN $7\frac{5}{8}$ 4/06			& 2.79						& 7.625					&	1690				&	2061				&	107				&	75.00		& 76.75					& -1.75						\\ \hline 
CPN $10\frac{1}{2}$ 5/06		& 2.82						& 10.50					&	1594				&	1813				&	-97				&	83.30		& 81.30					& 1.71						\\ \hline 
CPN $8\frac{3}{4}$ 7/07			& 3.77						& 8.75					&	1568				&	1807				&	68				&	74.52		& 74.52					& -1.34						\\ \hline 
CPN $7\frac{7}{8}$ 4/08			& 4.76						& 7.875					&	1376				&	1781				&	55				&	71.00		& 72.18					& -1.18						\\ \hline 
CPN $7\frac{3}{4}$ 4/09			& 5.79						& 7.75					&	1210				&	1610				&	0					&	71.00		& 71.00					& 0.00						\\ \hline 
CPN $8\frac{5}{8}$ 8/10			& 7.13						& 8.625					&	1110				&	1457				&	-13				&	73.50		& 73.15					& 0.35						\\ \hline 
CPN $8\frac{1}{2}$ 2/11			& 7.63						& 8.50					&	1023				&	1353				&	-71				&	75.00		& 72.93					& 2.07						\\ \hline 
\end{tabular}
	\caption{Calpine bonds as of 6/30/03}
	\label{tab:DAS}
\end{table}

Assuming an investor took a relative value trade in bonds maturing 4/06 and 5/06 as of the date in Table \ref{tab:DAS} in equal notional value (justified by closeness of maturities), such investor would have locked in \$3.4 per \$100 of notional value in such a trade, and would have realized at least that much in either bankruptcy (which actually occured in December 2005) or maturity, since in both cases the bonds would have paid off the same final amount.

\section{The CDS-Bond Basis} \label{sec:cds_bond_basis}

Although CDS and cash bonds reflect the same underlying issuer credit risk, there are important fundamental and technical reasons why the CDS and bond markets can sometimes diverge from the economic parity \cite{OKane-book}. Such divergences, commonly referred to as the CDS-Bond basis, are closely monitored by many credit investors. Trading the CDS-Bond basis is one of the widely used strategies for generation of excess returns using CDS. 

There are a number of both fundamental and technical reasons that affect the pricing of bonds and CDS and lead to the presence of the CDS-Bond basis even after correcting for the inherent biases associated with the commonly used Z-spreads or asset swap spreads. We list some of them below.

\noindent Factors that drive CDS spreads wider than bonds:
\begin{itemize}
\item	Delivery option: the standard CDS contract gives a protection buyer an option to choose a delivery instrument from a basket of deliverable securities in case of default. 
\item	Risk of technical default and restructuring: the standard CDS contract may be triggered by events that do not constitute a full default or a bankruptcy of the obligor. 
\item	Demand for protection: the difficulty of shorting credit risk in the bond market makes CDS a preferred alternative for hedgers and tends to push their spreads wider during the times of increasing credit risks.
\item	LIBOR-spread vs. Treasury spread: the CDS market implies trading relative to swaps curve, while most of the bond market trades relative to Treasury bond curve. Occasionally, the widening of the LIBOR spread that is driven by non-credit technical factors such as MBS hedging can make bonds appear {\em optically} tight to LIBOR.
\end{itemize}

\noindent Factors that drive CDS spreads tighter than bonds:
\begin{itemize}
\item	Implicit LIBOR-flat funding: the CDS spreads imply a LIBOR-flat funding rate, which makes them cheap from the perspective of many protection sellers, such as hedge funds and lower credit quality counterparties, who normally fund at higher rates.
\item	Counterparty credit risk: the protection buyer is exposed to the counterparty risk of the protection seller and must be compensated by tighter CDS spreads.
\item	Differential accrued interest loss: in the CDS market, the accrued interest is netted with the protection payment in case of default. In the bond market, the accrued coupon amount is often lost or added to outstanding notional which recovers only a fraction in default. 
\item	Differential liquidity: while the amount of the available liquidity in the top 50 or so bond issuers is greater in the cash market, the situation is often reverse for the rest of the credit market where writing protection can be easier than buying the bonds. 
\end{itemize}

For all these reasons, the CDS-Bond basis can be and often is substantial. While the fundamental factors affect the proper value of the fair basis and are largely stable, the transient nature of the more powerful technical factors causes the basis to fluctuate around this fair value with a typical mean reversion time that ranges between a few weeks to few months. This makes basis trading an attractive relative value investment strategy, albeit with its own inherent risks of liquidity-driven blowups, like any other such strategy. For example, during the market dislocation at the end of 2008 the CDS-Bond basis has reached in some cases several hundred basis points. 

Many investors have been actively trading such basis convergence strategies by relying on the conventional basis measure, the difference between the CDS spread and the bond's Z-spread. Since Z-spread itself is a biased measure of credit risk, therefore this conventional basis measure is also biased. In this section, we present the alternative measure, based on accurate replication of bonds with CDS.

\subsection{The CDS-Bond Complementarity} \label{sec:cds_bond_complementarity}

Assume that the underlying risk-free discount curve $r(t)$ (usually LIBOR) and the issuer's hazard rate term structure $h(t)$ are given. The forward base discount function and the forward survival probability are given by equations (\ref{eq:fwd-base-discount}) and (\ref{eq:fwd-survival-prob}), respectively.

Consider a credit-risky bond with a given coupon $C$ and final maturity $T$. The projected forward price of a fixed coupon bond depends on both riskless rate and hazard rate term structures as well as the level of the coupon in the following manner (for simplicity of exposition we use the continuous-time approximation and ignore the small corrections proportional to the coupon level - see Appendix A for detailed derivation):

\begin{eqnarray} \label{eq:bond-fwd-price}
	P(t,T) & = & C \, \int_{t}^{T}{ du \, e^{- \int_{t}^{u}{ ds \, \left(f(s)+h(s)\right)}}} + e^{- \int_{t}^{T} ds \, \left(f(s)+h(s)\right)}  \nonumber \\
	      & + & R \, \int_{t}^{T}{ du \, h(u) \, e^{- \int_{t}^{u}{ ds \, \left(f(s)+h(s)\right)}} }
\end{eqnarray}

The first term reflects the present value of the coupon stream under the condition that the bond survived until some intermediate time $u$, the second term reflects the present value of the final principal payment under the condition that the bond survived until the final maturity $T$, the third term reflects the recovery of the fraction $R$ of the face value if the issuer defaults at any time between the valuation time and the final maturity.
 
Let us define the {\em risk-free-equivalent coupon} stream $RFC(t,T)$ which would reproduce the same forward price term structure but only when discounted with the risk-free discount function, without any default probability. Such a coupon stream will not be constant in general and will have a non-trivial term structure, depending on both the underlying risk-free rates and, through the price of the risky bond, on the issuer hazard rates as well. The defining condition is:

\begin{equation} \label{eq:bond-rfc}
	P(t,T) = \int_{t}^{T}{ du \, RFC(t,u) \, e^{- \int_{t}^{u}{ ds \, \left(f(s)+h(s)\right)}}} + e^{- \int_{t}^{T}{ ds \, \left(f(s)+h(s)\right)}}  
\end{equation}

The concept of a risk-free equivalent coupon stream is necessary for consistent definition of the difference between the default-risky and risk-free bonds when the underlying interest and hazard rates have non-trivial term structures and the bonds are expected to deviate from par pricing either currently or at any time in the future.

To find the relationship between the risk-free-equivalent coupon stream $RFC(t,T)$ and the forward price $P(t,T)$ of a credit-risky bond, let us take a derivative with respect to the valuation time $t$ of both sides of equations (\ref{eq:bond-fwd-price}) and (\ref{eq:bond-rfc}). 

\begin{eqnarray} 
	\frac{\partial P(t,T)}{\partial t} & = & \left( r(t) + h(t)\right) \, P(t,T) - C - R \, h(t) \label{eq:bond-d-fwd-price} \\
	\frac{\partial P(t,T)}{\partial t} & = & r(t) \, P(t,T) - RFC(t,T) \label{eq:bond-d-rfc}
\end{eqnarray}

Since the left-hand sides are equal by construction, we can equate the right-hand sides and obtain the relationship between the risk-free-equivalent coupon stream and the forward price:

\begin{equation} \label{eq:rfc-bond-complementarity}
	C - RFC(t,T) = h(t) \left( P(t,T) - R \right)
\end{equation}

Consider now a forward CDS contract for the infinitesimal period $[t,t+dt]$. Since the hazard rate term structure can be ignored in such a short period, then the credit triangle formula (\ref{eq:CreditTriangleDerived}) applies and the forward CDS spread is simply proportional to the hazard rate for that period:

\begin{equation} \label{eq:cds-fwd-spread}
	S_{CDS}^{fwd}(t,t+dt) = \left( 1 - R \right) h(t) 
\end{equation}

Substituting this definition into equation (\ref{eq:rfc-bond-complementarity}), we get a complementarity condition between the credit-risky coupon, risk-free-equivalent coupon stream and the forward CDS spreads:

\begin{eqnarray} 
	RFC(t,T) & = & C - S_{CDS}^{fwd}(t,t+dt) \, N_{CDS}^{fwd}(t,T) \label{eq:cds-bond-complementarity} \\
	N_{CDS}^{fwd}(t,T) & = & \frac{P(t,T) - R}{1 - R}  \label{eq:cds-fwd-notional}
\end{eqnarray}
where we also introduced the forward CDS notional $N_{CDS}^{fwd}(t,T)$, which depends on the forward price of the credit bond.

\subsection{Static Hedging of Credit Bonds with CDS} \label{sec:cds_bond_static_hedging}

This relationship suggests a consistent hedging strategy for non-par credit-risky bonds which consists of a stream of forward CDS with notionals $N_{CDS}^{fwd}(t,T)$ depending on the forward price of the bond. The residual cashflows of the credit-risky bond after paying the required premiums coincide with the projected risk-free-equivalent coupon stream. Although there is still a timing risk associated with this hedging strategy, the notionals of the hedges are such that the recovered value will be equal to the correct forward price of the bond, and therefore the timing risk is unimportant when evaluating the present value of the hedged cashflows to the initial time or to any future time before maturity. 

Alternatively, one could replace the forward CDS in this hedging strategy with pairs of spot CDS of increasing maturities. If one chose a grid of maturities $t_{i}$ for which the hedging is done, the corresponding notionals of the pairs of CDS for each interval $[t_{i},t_{i+1}]$ would be given by:

\begin{equation} \label{eq:spot-cds-hedge}
	N_{CDS}^{spot}(t_{i},t_{i+1},T) = \frac{P(t_{i},T) - P(t_{i+1},T)}{1 - R}
\end{equation}

If we execute the hedging strategy with long-short pairs, the result becomes a staggered hedge which is nearly 100\% notional for the final maturity, and which includes some additional relatively small long (or short) positions for shorter maturities depending on the forward prices of the credit bond being hedged. Each such position hedges the incremental digital price risk (with no recovery) corresponding to the next maturity interval on the hedging grid.

\begin{figure}[t]
\includegraphics[height=3in,width=4.5in]{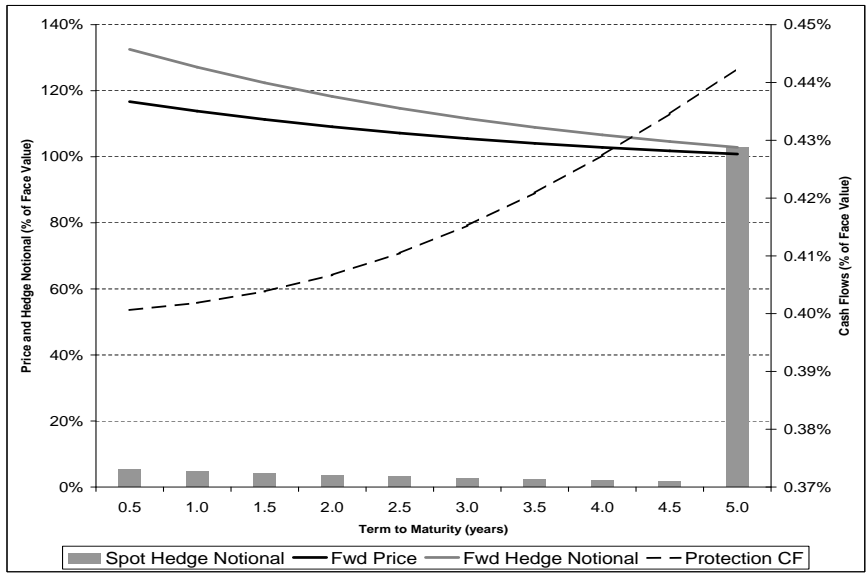}%
\caption{{\small Static hedging of a premium bond by CDS.}}%
\label{fig:Bond-CDS-basis-1}%
\end{figure}

\begin{figure}[t]
\includegraphics[height=3in,width=4.5in]{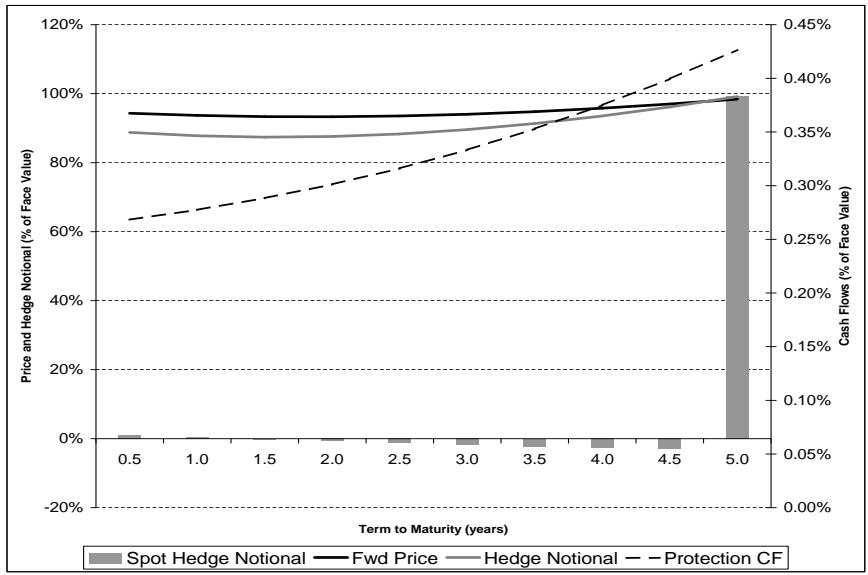}%
\caption{{\small Static hedging of a discount bond by CDS.}}%
\label{fig:Bond-CDS-basis-2}%
\end{figure}

\begin{figure}[t]
\includegraphics[height=3in,width=4.5in]{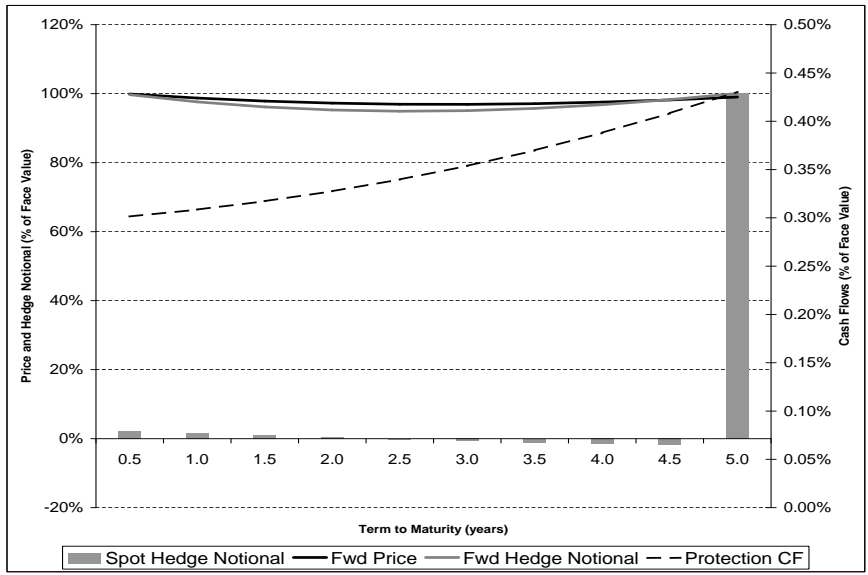}%
\caption{{\small Static hedging of a near-par bond by CDS.}}%
\label{fig:Bond-CDS-basis-3}%
\end{figure}

From the earlier discussions in this chapter, it is clear that both the coupon level of the credit bond and the term structure of the underlying interest rates and issuer's hazard rates may substantially affect the hedging strategy with forward CDS or long-short CDS pairs. Its dependence on the underlying bond is depicted in Figures \ref{fig:Bond-CDS-basis-1}, \ref{fig:Bond-CDS-basis-2} and \ref{fig:Bond-CDS-basis-3}. 

Figure \ref{fig:Bond-CDS-basis-1} shows a case of a bond with a high coupon equal to 8\% and a high current price of 116.69\%. The forward price of the credit bond gradually declines toward maturity. The forward CDS hedge notional mirrors that behavior, starting as high as 133\% of the face value, and gradually decreases toward 100\%. The spot CDS hedge contains the final notional hedge and relatively large additional hedges at earlier maturities. Despite the decrease in the forward hedge notional, the semi-annual cost of hedging grows gradually from 40bp to 44bp as a result of a relatively steep forward CDS curve term structure. 

Figure \ref{fig:Bond-CDS-basis-2} shows a case of a bond with a very low coupon equal to 3\% and a current discount price equal to 94.33\%. The forward price of the bond gradually grows towards maturity. The forward CDS hedge notional mirrors this, starting at 89\% of the face value, and gradually increases toward 100\%. The spot CDS hedge contains the final notional hedge and additional hedges at intermediate maturities which actually change sign, with some offsetting short protection positions before the final maturity. The semi-annual cost of hedging grows more steeply from 27bp to 43bp as both forward CDS rates and hedge notionals grow. 

Figure \ref{fig:Bond-CDS-basis-3} shows a case of a bond with near-par coupon equal to 4.25\% and a current price of 99.93\%. Despite the fact that this is a par bond, the forward bond prices and forward CDS hedge notionals exhibit a non-trivial term structure, starting near 100\%, then dropping to lower levels and only pulling back to par near final maturity. The spot CDS hedge also has non-trivial intermediate maturity hedges with changing sign. The semi-annual cost of hedging grows from 33bp to 43bp, which is somewhere between the high and low coupon cases. 

The static hedging strategy using CDS addresses only the credit risk exposure. According to the complementarity principle proven in (\ref{eq:cds-bond-complementarity}), the remainder of these hedging strategies is the risk-free-coupon stream (RFC) bond whose forward price profile matches precisely that of the credit bond. While this hypothetical RFC bond is no longer subject to credit loss risk, it is still subject to interest rate risk. In order to fully hedge the residual interest rate risk one would simply have to swap all the projected RFC cash flows into floating rate using a sequence of interest rate swaps of appropriate maturities. 

This, however, does not fully eliminate the interest rate risk. Indeed, upon a default event the cash flows from the credit bond itself and all the CDS will terminate, while the interest rate swaps that have maturities longer than the date of default will still be outstanding. The expected net market value of these remaining swaps is equal to the expected variation of the interest rate hedge package from the forward price of the RFC bond. Hence, it is equal to zero by construction regardless of the default timing. In order to fully eliminate not only the expected risk exposure but also the residual risk in all states of the world both before and after default event one would need to use the so-called {\em extinguishing asset swaps}, i.e. fixed-for-floating swaps which contractually terminate upon the default of the reference credit entity.

\subsection{Consistent Measures for CDS-Bond Basis} \label{sec:cds_bond_basis_measures}

If the bonds of a given issuer were perfectly priced according to our framework, and if the term structure of credit risk as well as the recovery value was in agreement between the bond and CDS markets, the bond prices would satisfy the survival-based fair value given by (\ref{eq:pricing-FRP-simple}). The hazard rate in this equation would be identical to the one calibrated from the credit triangle relationship (\ref{eq:CalibrateDDS}), and the BCDS term structure (\ref{eq:BCDS}) would coincide with market observed CDS term structure. There would be no basis between the two markets, and the hedged cash bond would have zero expected excess return over the appropriately defined risk-free rate. 

If, on the contrary, the market-observed CDS spreads were tighter than BCDS, one could hedge the credit bond by CDS at a lower cost than that implied by the model, thus locking in a positive expected excess return. Vice versa, if the market CDS spreads were wider than BCDS, the hedge would have a higher cost, and the expected excess return of the hedged position would be negative. Such non-zero expected excess return is what is supposed to be reflected by the consistent measure of the Bond-CDS basis.

When the two markets, bonds and CDS, show different levels of implied default probability, there is an ambiguity as to which of these is correct, if any. It is a common assumption that if there exists a liquid CDS market with full quoted term structure, then it is this market which is less biased as far as credit risk is concerned. We share this opinion since the CDS market is naturally focused on this particular source of risk, and since the metrics used (CDS spread, risky PV01, etc.) by most market participants are unbiased, unlike the case of cash bonds where most market participants use biased metrics like Z-spread.

In this case, the natural definition of the CDS-Bond basis would result from the comparison of the bond market price with its fair value in the survival-based framework where the hazard rate is taken from the CDS market. In the same way as the default-adjusted spread reflects the pricing basis of a given bond to best fit of the issuer or sector, the additional discount spread in this equation can be regarded as the pricing basis between this particular bond and the CDS market. Thus, the definition of the Basis Spread (BS) is:

\begin{eqnarray} \label{eq:cds-bond-basis}
	P_{mkt} & = & Z_{base}(t_{N}) \, Q_{CDS}(t_{N}) \, e^{-BS \, t_N}  \nonumber \\
	      & + & \frac{C}{q} \sum_{i=1}^{N}{ Z_{base}(t_{i}) \, Q_{CDS}(t_{i})  \, e^{-BS \, t_i}} \nonumber \\
	      & + & R \, \left(1 + \frac{C}{2q}\right) \, \sum_{i}^{N}{ Z_{base}(t_{i}) \, \left(Q_{CDS}(t_{i-1}) - Q_{CDS}(t_{i})\right)  \, e^{-BS \, t_i} } \nonumber \\
	      & - & C \, t_{acc} 
\end{eqnarray}

In the opposite case, if the CDS market is not very liquid beyond the benchmark 5-year maturity while there are many cash bonds allowing one to derive a well-defined term structure of credit risk on the bond side, it would be natural to consider the latter as the unbiased measure of credit risk. Potentially, one can also take a compromise view and consider the bond market as the correct source for the shape of the credit risk term structure, but recalibrate the level of this risk by fitting the 5-year CDS. One would then use this survuval term structure in (\ref{eq:cds-bond-basis}) for determining the basis for a particular bond.


\subsection{The Coarse-Grained Hedging and Approximate Basis} \label{sec:coarse-grained-basis}

The static hedging strategy using a sequence of forward CDS is difficult to implement in practice. Although less precise, the staggered strategy using spot CDS is generally easier to put to work. However, even the staggered strategy, if implemented in short term increments as presented in Figures \ref{fig:Bond-CDS-basis-1}-\ref{fig:Bond-CDS-basis-3}, would lead to odd-lot hedge notionals for intermediate terms, and likely result in an unacceptable loss of liquidity. 

As a compromise between accuracy and liquidity, we suggest a coarse-grained staggered hedge which can be constructed using a maturity grid with longer intervals. The forward price changes in these intervals will yield lumpier intermediate hedge notionals, according to (\ref{eq:spot-cds-hedge}). The optimal hedging grid will depend on the bond coupon level and the underlying interest rates. For bonds with modest premium or discount, just one or two additional hedges can result in sufficient accuracy.

\begin{figure}[t]
\includegraphics[height=3in,width=4.5in]{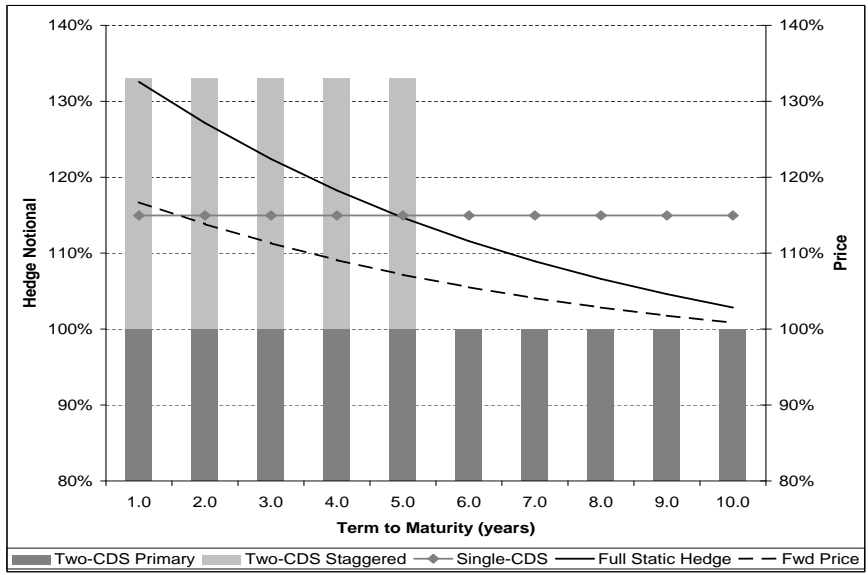}%
\caption{{\small Coarse-grained hedging of a premium bond by CDS.}}%
\label{fig:Bond-CDS-basis-4}%
\end{figure}

Let us consider a single-CDS strategy first. Figure \ref{fig:Bond-CDS-basis-4} shows an example of such strategy (the line with diamonds) contrasted with the theoretical precise strategy using forwards (the light solid line) in the case of a premium credit bond whose forward price (dashed line) gradually approaches par towards the 5-year maturity. Such a simple hedging strategy would typically be underhedged (compared to the theoretical requirement) during the early years and overhedged during the later years of its projected existence. The optimal hedge notional will the such that the net present value of the outstanding risk exposures (positive for short terms and negative for long terms) become precisely zero. 

Such a requirement also helps in explaining why the `market hedge ratio' that is equal simply to the price premium of the underlying bond (which is quite popular among the practitioners, see \cite{McAdie-OKane-2001}) provides a good starting guess for the correct hedge amount. Indeed, had we approximated the theoretical hedge notional line by a straight line, and ignored the effect of interest rate discounting, the optimal notional would correspond to the mid-point between the final theoretical hedge ratio at maturity, i.e. 100\%, and the initial theoretical hedge ratio at current time (\ref{eq:cds-fwd-spread}). If one used a recovery rate of $R=50\%$ which is quite close to the long-term average recovery estimates, one would obtain the hedge notional equal to the current price of the bond. 

We can see therefore, that the market practice is not too different from the correct single-CDS optimal hedge. One must ask, however, whether the single-CDS hedge itself is the optimal solution, or are we missing something important by limiting ourselves to only one hedging instrument.

Consider now the two-CDS hedging strategy depicted in Figure \ref{fig:Bond-CDS-basis-4} by stacked bars. The strategy consists of a face value hedge to final maturity, plus an additional (staggered) hedge to a shorter maturity. We can arrange this strategy to be, for example, overhedged during the earlier years and underhedged during the later years of its projected existence. The same requirement of zero net present value of the outstanding risk exposures will define the optimal notional of the smaller staggered hedge, given its chosen maturity. 

Note also, that by allowing the maturity of the add-on staggered hedge to be equal to the final maturity of the bond we would recover the case of a single-CDS strategy. Therefore, the two-CDS strategy will always be at least as good as the single-CDS one. When will it be better? It would be a better choice when the cost of hedging using two CDS turns out to be less than the cost of hedging using a single CDS. In the case of a premium bond, this will be the case if the CDS spread term structure is upward sloping, making the use of shorter term CDS a preferable option.

The considerations above lead us to the following simple recipe for a practical and accurate hedging strategy of credit bonds with CDS:
\begin{itemize}
\item Set the primary hedge to the final maturity of the bond with a notional equal to par face value of the bond.
\item Consider various intermediate maturities (including the bond's final maturity) for which a sufficiently liquid market in CDS exists, and determine the optimal staggered hedge amount for the second CDS position for each maturity.
\item Select the lowest cost hedging strategy among all considered two-CDS combinations.
\end{itemize}

The corresponding approximate CDS-Bond basis for the bond under consideration will be determined by comparison of its theoretical full excess spread (\ref{eq:bond-excess-spread}) where the P-spread is measured with respect to LIBOR base curve, and the rpv01-weighted aggregate CDS spread of the coarse-grained staggered hedging strategy. This means that, even in a simplified framework, the correct relative value measure for basis trading strongly depends on the bond's coupon level and price premium, and the term structures of both interest rates and CDS spreads.


\section{Conclusions}

In this chapter we have critically examined the conventional bond pricing methodology and have shown that it does not adequately reflect the nature of the credit risk faced by investors. In particular, we have demonstrated that the strippable discounted cash flows valuation assumption which is normally taken for granted by most analysts, leads to biased estimates of relative value for credit bonds. Moreover, even the CDS market which does not use this pricing methodology is strongly influenced by its prevalence in the cash market because of the constant cross-benchmarking of cash bonds and CDS. The example of the `optically distorted' term structures of Z-spreads and CDS spreads of such a highly liquid name as Ford Motor Credit shown in section \ref{sec:phenomenology} should convince investors that the conventional methodology can indeed be quite misleading.

We have introduced a consistent survival-based valuation methodology which is free of the biases mentioned above, albeit at a price of abandoning of the strippable discounted cash flows valuation assumption. We also developed a robust estimation methodology for survival probability term structures using the exponential splines approximation, and implemented and tested this methodology in a wide variety of market conditions and across a large set of sectors and issuers, from the highest credit quality to highly distressed ones. 

To remedy the loss of intuition due to the abandonment of OAS, Z-spreads, and other conventional spread measures, we have introduced a host of new definitions for credit term structures, ranging from valuation measures such as the hazard rate and P-spread to relative value measures such as the bond-specific Default Adjusted Spread. 

In conclusion, we believe that the adoption of the survival-based methodologies advocated in this chapter by market participants will lead to an increase in the efficiency of the credit markets just as the adoption of better prepayment models led to efficiency in the MBS markets twenty years ago. Investors who will be at the forefront of this change will be in a position to benefit from the secular shift to a much more quantitative approach to credit portfolio management. We have witnessed many such turning points in recent years, including the widespread following attained by structural credit risk models pioneered by Merton and further developed by KMV and others, the explosive growth in the credit derivatives and structured credit market which proceeded alongside with a dramatic progress in modelling complex risks of correlated defaults and losses, the adoption of new banking regulatory standards which place a much greater emphasis on quantitative measures of credit and counterparty risks, and finally the proliferation of credit hedge funds and relative value investors who stand ready to exploit any inefficiencies still present in the marketplace. It was long overdue that the most traditional of all credit instruments, the credit bonds, would also be considered in the light of this new knowledge. 

Acknowledgements: I would like to thank my collaborators Roy Mashal and Peili Wang with whom much of the presented methodology was implemented while the author was at Lehman Brothers, as well as Marco Naldi, Mark Howard and many other colleagues at (formerly) Lehman Brothers Fixed Income Research department for numerous helpful discussions.

\appendix

\section{Exponential Splines} \label{sec:splines}

The exponential spline approximation is defined in a way which facilitates smooth fitting of functions which are exponentially decreasing with term to maturity but are not necessarily required to have a constant rate of decrease. The shape of the approximated function is given by a linear combination of spline factors:

\begin{equation} \label{eq:exp-spline}
F(t) = \sum_{k=1}^{K}{\beta_{k} \, \Phi_{k}\left(t\left|\eta \right. \right)}
\end{equation}
where the spline coefficients  $\beta_{k}$ are constants derived from the best fit optimization procedure, such as minimization of the weighted average square of bond pricing errors in the case of fitting Treasury discount functions or credit survival probability term structures. 

The exponential spline functions $\Phi_{k}\left(t\left|\eta \right. \right)$ have a fixed shape depending only on the remaining term to maturity, and on the decay parameter $\eta$. 

The first three spline factors are known as no-knot factors because they are smooth in the entire range of maturities. 

\begin{equation} \label{eq:exp-spline-no-knot}
\Phi_{k}\left(t\left|\eta \right. \right) = e^{-k \, \eta \, t}
\end{equation}

In fixed income applications, the shape of the yield curve often reflects market segmentation - the short, medium and long maturities can have substantially different behavior. To address this, one uses higher order spline factors (number 4 and above) which are exactly zero below certain maturity known as the {\em knot point} $T_{k}^{knot}$, have the familiar exponential shape above the knot point, and have a smooth value and first derivative at the knot point itself. These requirements determine the higher order spline factors uniquely as follows:

\begin{equation} \label{eq:exp-spline-knot}
\Phi_{k}\left(t\left|\eta \right. \right) = \Theta\left(t- T_{k}^{knot}\right) \, \left(\frac{1}{3} - e^{- \eta \, (t-T_{k}^{knot})} + e^{- 2 \, \eta \, (t-T_{k}^{knot})} + \frac{1}{3} e^{- 3 \, \eta \, (t-T_{k}^{knot})} \right)
\end{equation}

\begin{figure}[t]
\includegraphics[height=3in,width=4.5in]{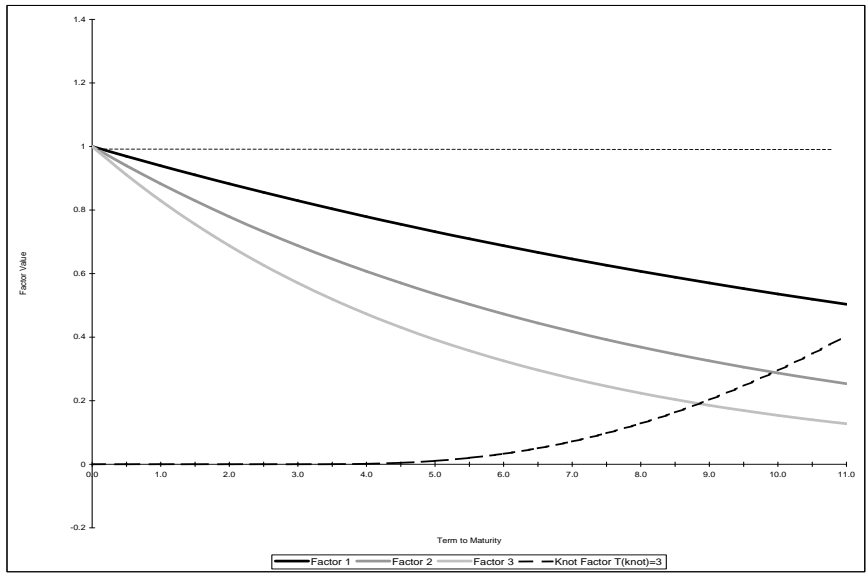}%
\caption{{\small Exponential spline factors.}}%
\label{fig:exp-splines}%
\end{figure}

The shapes of the spline factors are shown in Figure \ref{fig:exp-splines}.

\section{Continuous Time Approximation for Credit Bond and CDS Pricing} \label{sec:continuous_time}

Continuous compounding is a convenient technique which may often simplify the analysis of relative value and forward pricing of credit bonds. It corresponds to coupon payments being made continuously. The present value for a hypothetical continuously compounded credit-risky bond can be calculated using the instantaneous forward interest rates (\ref{eq:fwd_rate}) and hazard rates (\ref{eq:ts-hazard-spline}). The base discount function and the survival probability are given by (\ref{eq:fwd-base-discount}) and (\ref{eq:fwd-survival-prob}), respectively:


Assuming uncorrelated interest, hazard and recovery rates, one can combine equations (\ref{eq:fwd-base-discount}) and (\ref{eq:fwd-survival-prob}) to obtain a continuously compounded analog of the bond pricing equation (\ref{eq:pricing-FRP-simple}):

\begin{eqnarray} \label{eq:bond-price-ccomp}
	P(T) & = & C \, \int_{0}^{T}{ du \, e^{- \int_{0}^{u}{ ds \, \left(f(s)+h(s)\right)}}} + e^{- \int_{0}^{T} ds \, \left(f(s)+h(s)\right)}  \nonumber \\
	      & + & R \, \int_{0}^{T}{ du \, h(u) \, e^{- \int_{t}^{u}{ ds \, \left(f(s)+h(s)\right)}} }
\end{eqnarray}

This simple formula overestimates the present value of a credit bond for two distinct reasons:
\begin{itemize}
\item	First, it neglects the expected accrued coupon loss and recovery in case of default. 
\item	Second, it overestimates the present value of the regular coupon payments because it presumes that portions of the coupon were paid earlier and it discounts those portions with a correspondingly smaller discount factor (and higher survival probability).
\end{itemize}

The more accurate approximation which we derive here corrects for these two biases.

The correction for the coupon loss bias can be estimated by noting that the expected timing of the default event under a constant hazard rate assumption is roughly half-way through the payment period. Hence, the expected accrued interest equals to approximately half of the scheduled coupon payment, which in turn is equal to $1/q$ fraction of the coupon rate for a bond with frequency $q$. Consequently, to correct for this bias we should explicitly add the expected accrued interest at default to the principal amount, assuming that the coupon and principal recovery are the same $R_{int}=R_{pr}=R$.

The correction for the early discount bias can be estimated by noting that by distributing the coupon payments evenly between the two coupon dates we get the survival-weighted present value which is roughly half-way between the present value of a bullet coupon payment on the two ends of the coupon period. Thus, for each bullet coupon the continuous-time formula (\ref{eq:bond-price-ccomp}) corresponds to a present value bias equal to half of difference between the `true' present value of the earlier coupon payment and the current coupon payment. When summing up all of these biases, the corrections for all intermediate coupon payments cancel each other, and the total present value bias is simply half of the difference between the present value of the first coupon payment and the last coupon payment. For the valuation date just prior to a coupon payment, this results in a simple estimate since the present value of that impending coupon payment is simply equal to its amount. For other valuation dates the situation is slightly more complicated but the approximation remains pretty close nevertheless.  

We obtain the continuous-time approximation for the clean price of an $q$-frequency credit bond by subtracting these two bias estimates from the original `naïve' formula. 
Finally, we should also include the Default-Adjusted Spread (DAS), an issue-specific discounting measure introduced in Part 1 of this series that allows us to use the issuer- or sector-specific hazard rate term structure while exactly fitting the observed price of individual bonds. 

The final formula for the clean price of a fixed-coupon credit bond in the continuous-time approximation is:

\begin{eqnarray} \label{eq:bond-price-ccomp-correct}
	P\left(T\left|q\right.\right) & = & C \, \int_{0}^{T}{ du \, e^{- \int_{0}^{u}{ ds \, \left(f(s)+h(s)+DAS\right)}}} + e^{- \int_{0}^{T} ds \, \left(f(s)+h(s)+DAS\right)}  \nonumber \\
	      & - & \frac{C}{2q} \, \left(1 - e^{- \int_{0}^{T} ds \, \left(f(s)+h(s)+DAS\right)} \right)  \nonumber \\
	      & + & R \, \left(1 + \frac{C}{2q} \right) \, \int_{0}^{T}{ du \, h(u) \, e^{- \int_{t}^{u}{ ds \, \left(f(s)+h(s)+DAS\right)}} }
\end{eqnarray}

This approximation is quite accurate across all values of coupons and for all shapes and levels of the underlying interest rate and hazard rate curves. Both correction terms can be quite important.  

In a similar fashion, we can write the pricing equation for CDS in continuous compounding approximation as:

\begin{eqnarray} \label{eq:CDS-pricing-ccomp}
	 UP(T) & + & C_{CDS} \, \int_{0}^{T}{ du \, e^{- \int_{0}^{u}{ ds \, \left(f(s)+h(s)\right)}}}  \nonumber \\
	      & = & (1-R) \, \int_{0}^{T}{ du \, h(u) \, e^{- \int_{t}^{u}{ ds \, \left(f(s)+h(s)\right)}} }
\end{eqnarray}

After correcting for the biases associated with finite frequency (typically, $q=4$) of premium payments, namely the discounting bias and the netting convention between unpaid accrued premium and the protection payment, we get:

\begin{eqnarray} \label{eq:CDS-pricing-ccomp-correct}
	 UP\left(T\left|q\right.\right) & + & C_{CDS} \, \int_{0}^{T}{ du \, e^{- \int_{0}^{u}{ ds \, \left(f(s)+h(s)\right)}}}  \nonumber \\
	      & - & \frac{C_{CDS}}{2q} \, \left(1 - e^{- \int_{0}^{T}{ ds \, \left(f(s)+h(s)\right)}} \right)  \nonumber \\
	      & = & \left(1 - R - \frac{C_{CDS}}{2q} \right) \, \int_{0}^{T}{ du \, h(u) \, e^{- \int_{t}^{u}{ ds \, \left(f(s)+h(s)\right)}} }
\end{eqnarray}

Finally, the continuous-time approximation for the par CDS spread is given by the solving for $C_{CDS}$ in the above equation for the case $UP\left(T\left|q\right.\right) = 0$:

\begin{equation} \label{eq:bcds-ccomp}
	S_{CDS}\left(T\left|q\right.\right) = (1-R) \frac{\int_{0}^{T}{ du \, h(u) \, e^{- \int_{t}^{u}{ ds \, \left(f(s)+h(s)\right)}}}}{\int_{0}^{T}{ du \, \left(1 -  \frac{f(u)}{2q} \right) \, e^{- \int_{0}^{u}{ ds \, \left(f(s)+h(s)\right)}}}}
\end{equation}

In particular, we get a simple correction to the credit triangle formula for the case of flat term structures of interest and hazard rates:

\begin{equation} \label{eq:bcds-ccomp-flat}
	S_{CDS}\left(T\left|q\right.\right) = \frac{1}{1 -  \frac{f}{2q}} \, (1-R) \, h 
\end{equation}

\end{document}